\documentclass[aip,graphicx,reprint]{revtex4-1}
\usepackage{graphicx}
\usepackage{dcolumn}
\usepackage{bm}
\usepackage{color}

\draft 

\begin{document}





\title{Amino-acid-dependent main-chain torsion-energy terms
for protein systems} 

\author{Yoshitake Sakae}
\affiliation{Department of Theoretical and Computational Molecular Science, 
Institute for Molecular Science, Okazaki, Aichi 444-8585, Japan}
\affiliation{Department of Physics, Graduate School of Science, 
Nagoya University, Nagoya, Aichi 464-8602, Japan}
\author{Yuko Okamoto}
\affiliation{Department of Physics, Graduate School of Science, 
Nagoya University, Nagoya, Aichi 464-8602, Japan}
\affiliation{Structural Biology Research Center, Graduate School of
Science, Nagoya University, Nagoya, Aichi 464-8602, Japan}
\affiliation{Center for Computational Science, Graduate School of
Engineering, Nagoya University, Nagoya, Aichi 464-8603, Japan}


















\date{\today}

\begin{abstract}
Many commonly used force fields for protein systems such as AMBER, CHARMM, 
GROMACS, OPLS, and ECEPP have  
amino-acid-independent force-field parameters of main-chain 
torsion-energy terms.
Here, we propose a new type of amino-acid-dependent torsion-energy terms
in the force fields.
As an example, we applied this approach to AMBER ff03 force field 
and determined new amino-acid-dependent parameters for $\psi$ 
and $\psi'$ angles
for each amino acid by using our optimization method, 
which is one of the knowledge-based approach.
In order to test the validity of the new force-field parameters,
we then performed folding simulations of $\alpha$-helical and 
$\beta$-hairpin peptides, 
using the optimized force field. 
The results showed that the new force-field parameters gave structures 
more consistent with the 
experimental implications than the original AMBER ff03 force field.
\end{abstract}


\maketitle 



\section{Introduction}
Computer simulations of protein folding into native structures
can be achieved when both of the following two requirements are met:
(1) potential energy functions (or, force fields) for the protein
systems are sufficiently accurate
and (2) sufficiently powerful conformational sampling methods are
available.
Professor Harold A. Scheraga has been one of the most important
pioneers in studies of both of the above 
requirements \cite{ScheragaRev1,Scheragarev2}.
By the developments of the generalized-ensemble algorithms
(for reviews, see, e.g., Refs.~\cite{HOrev,GEA}) and related
methods, Requirement (2) seems to be almost fulfilled.
In this article, we therefore concentrate our attention on 
Requirement (1). 

There are several well-known all-atom (or united-atom) force fields, 
such as AMBER \cite{parm94, parm96_2, parm99, parm99SB, parm03}, 
CHARMM \cite{charmm,CMAP}, OPLS \cite{opls1,opls2}, 
GROMOS \cite{gromos_v2,gromos_v3}, 
GROMACS \cite{gromacs1,gromacs2}, 
and ECEPP \cite{ECEPP,ECEPP05}.
Generally, the force-field parameters are determined based on experimental results for small molecules 
and theoretical results using quantum chemistry calculations of small peptides such as alanine dipeptide.

In a force field, the potential energy is usually composed of the bond-stretching term, the bond-bending term, 
the torsion-energy term, and the nonbonded energy term.
In these energy terms, it is known that the torsion-energy term is the most problematic.
For instance, the ff94 \cite{parm94} and ff96 \cite{parm96} versions of AMBER differ only in 
the main-chain torsion-energy parameters.
Nevertheless, the secondary-structure-forming tendencies of the two force fields 
are quite different \cite{YSO1,YSO2, SO1, SO2, SO3}. 
Therefore, many researchers have studied this main-chain torsion-energy terms and their force-field parameters.
For instance, newer force-field parameters of the main-chain torsion-energy terms about $\phi$ and $\psi$ angles 
have been developed, which are, e.g., AMBER ff99SB \cite{parm99SB}, AMBER ff03 \cite{parm03}, 
CHARMM22/CMAP \cite{CMAP} and OPLS-AA/L \cite{opls2}.
The methods of the force-field refinement thus mainly concentrate on the torsion-energy terms.
These modifications of the torsion energy are usually based on 
quantum chemistry calculations \cite{Carlos,Duan,IWA,MFB,Kamiya} or NMR experimental results \cite{Hummer2009, Best_folding}.

We have also proposed a new main-chain torsion-energy term, which is 
represented by a double Fourier series in two variables,
the main-chain dihedral angles $\phi$ and $\psi$ \cite{SO4,SO5}.
This expression gives a natural representation of the torsion energy in the Ramachandran space 
\cite{Rama_Sasi}
in the sense that any two-dimensional energy surface periodic in both $\phi$ and $\psi$ can be
expanded by the double Fourier series.
We can then easily control secondary-structure-forming tendencies by modifying 
the main-chain torsion-energy surface.
We have presented preliminary results for AMBER ff94 and AMBER ff96 \cite{SO4,SO5}.
Moreover, we have introduced several optimization methods of force-field parameters \cite{SO1,SO2,SO3,SO6,SO7}.
These methods are based on the minimization of some score functions by 
simulations in the force-field parameter space,
where the score functions are derived from the protein coordinate data in the Protein Data Bank (PDB).
One of the score functions consists of the sum of the square of the force acting on each atom 
in the proteins with the structures from the PDB \cite{SO1,SO2,SO3}. 
Other score functions are taken from the root-mean-square deviations between the original PDB structures and 
the corresponding minimized structures \cite{SO6,SO7}.

In this article, we propose a new type of the main-chain torsion-energy 
terms for protein systems, 
which can have amino-acid-dependent force-field parameters.
As an example of this formulation, we applied this approach to 
the AMBER ff03 force field 
and determined new amino-acid-dependent main-chain torsion-energy parameters 
for $\psi$ (N-C$_{\alpha}$-C-N) and $\psi'$ (C$_{\beta}$-C$_{\alpha}$-C-N)
by using our optimization method in Refs \cite{SO1,SO2,SO3}.

In section 2 the details of the new main-chain torsion-energy terms are given.
In section 3 the results of applications of the method to AMBER ff03 force field and those of 
folding simulations of two peptides are presented. 
Section 4 is devoted to conclusions.

\section{Methods}
\subsection{Amino-acid-dependent force-field parameters}
The existing force fields for protein systems such as AMBER \cite{parm94, parm96_2, parm99, parm99SB, parm03}, CHARMM \cite{charmm,CMAP}, 
and OPLS \cite{opls1,opls2}, etc. use essentially the same functional forms 
for the potential energy $E_{\rm conf}$ except for minor differences.
The conformational potential energy $E_{\rm conf}$ can be written as, for instance,
\begin{equation}
E_{\rm conf} = E_{\rm BL} + E_{\rm BA} + E_{\rm torsion} + E_{\rm nonbond}~.
\label{ene_conf}
\end{equation}
Here, $E_{\rm BL}$, $E_{\rm BA}$, $E_{\rm torsion}$, and $E_{\rm nonbond}$
represent the bond-stretching term, the 
bond-bending term, the torsion-energy term, and the nonbonded energy term, respectively.
Each force field has similar but slightly different parameter values.
For example, the torsion-energy term is usually given by 
\begin{equation}
E_{\rm torsion} = \sum_{\Phi} \sum_n \frac{V_n\left(\Phi\right)}{2} 
\left\{ 1 + \cos \left[n \Phi - \gamma_n\left(\Phi\right)\right] \right\}~, 
\label{ene_torsion1}
\end{equation}
where the first summation 
is taken over all dihedral angles $\Phi$ (both in 
the main chain and in the side chains), 
$n$ is the number of waves, $\gamma_n$ is the phase, and $V_n$ is 
the Fourier coefficient.
Namely, the energy term $E_{\rm torsion}$ has $\gamma_n (\Phi)$ 
and $V_n(\Phi)$ as 
force-field parameters.

We can further write the torsion-energy term as
\begin{equation}
E_{\rm torsion} = E_{\rm torsion}^{(\rm MC)} + E_{\rm torsion}^{(\rm SC)}~,
\label{ene_torsion2}
\end{equation}
where
$E_{\rm torsion}^{(\rm MC)}$ and $E_{\rm torsion}^{(\rm SC)}$ are
the torsion-energy terms for dihedral angles around
main-chain bonds and around side-chain bonds, respectively.
Examples of the dihedral angles in
$E_{\rm torsion}^{(\rm MC)}$ are
$\phi$ (C-N-C$_{\alpha}$-C), 
$\psi$ (N-C$_{\alpha}$-C-N), $\phi'$ (C$_{\beta}$-C$_{\alpha}$-N-C), 
$\psi'$ (C$_{\beta}$-C$_{\alpha}$-C-N), and 
$\omega$ (C$_{\alpha}$-C-N-C$_{\alpha}$).
The force-field parameters in
$E_{\rm torsion}^{(\rm SC)}$ can readily depend on amino-acid
residues.
However, those in
$E_{\rm torsion}^{(\rm MC)}$ are usually taken to be
independent of amino-acid residues and the common parameter values
are used for all the amino-acid residues (except for proline).
This is because the amino-acid dependence of the force field
is believed to be taken care of by the very existence of side
chains.  
In Table I, we list examples of the
parameter values for
$\psi$ (N-C$_{\alpha}$-C-N)  and $\psi'$ (C$_{\beta}$-C$_{\alpha}$-C-N) 
in general AMBER force fields.
    
However, this amino-acid independence of the main-chain 
torsion-energy terms is not an absolute requirement, because
we are representing the entire force field by rather a small
number of classical-mechanical terms.  In order to reproduce
the exact quantum-mechanical contributions, one can introduce
amino-acid dependence on any force-field term including
the main-chain torsion-energy terms.
Hence, we can generalize 
$E_{\rm torsion}^{(\rm MC)}$ in Eq.~(\ref{ene_torsion2})  
from the expression in Eq.~(\ref{ene_torsion1})
to the following amino-acid-dependent form:
\begin{widetext}
\begin{equation}
E_{\rm torsion}^{(\rm MC)} = \sum_{k=1}^{20} 
\sum_{\Phi_{\rm MC}^{(k)}} 
\sum_n \frac{V_n\left(\Phi_{\rm MC}^{(k)}\right)}{2} 
\left\{ 1 + \cos \left[n \Phi_{\rm MC}^{(k)} - 
\gamma_n\left(\Phi_{\rm MC}^{(k)}\right)\right] \right\}~, 
\label{ene_torsion3}
\end{equation}
\end{widetext}
where $k$ ($=1, 2, \cdots, 20$) is the label for the 20 
kinds of amino-acid residues and 
$\Phi_{\rm MC}^{(k)}$ are dihedral angles around the
main-chain bonds in the $k$-th amino-acid
residue.


\subsection{Optimization method for force-field parameters}
In the previous subsection, we have generalized the main-chain
torsion-energy term
$E_{\rm torsion}^{(\rm MC)}$ so that its parameters are
amino-acid dependent.  The question is then how to obtain
optimal parameter values for this new main-chain torsion-energy
term.

One method is to use the parameter optimization method that
was introduced in Refs.~\cite{SO1,SO2,SO3}.
We first retrieve $N$ native structures (one structure per protein) from PDB.
We try to choose proteins from different amino-acid sequence homology as much as possible.
If the force-field parameters are of ideal values, then
all the chosen native structures are stable without any force acting on
each atom in the molecules on the average.  Hence, we expect
\begin{equation}
F = 0~,
\label{F0}
\end{equation}
where
\begin{equation}
F = \sum_{m=1}^N \frac{1}{N_m} \sum_{i_m = 1}^{N_m} \left| \vec{f}_{i_m} \right| ^2~,
\label{F}
\end{equation}
and
\begin{equation}
\vec{f}_{i_m} = - \frac{\partial E_{\rm total}^{\{m\}}}{\partial \vec{x}_{i_m}}~. 
\label{F1}
\end{equation}
Here, $N_m$ is the total number of atoms in molecule $m$, $E_{\rm total}^{\{m\}}$ is
the total potential energy for molecule $m$, and $\vec{f}_{i}$ is 
the force acting on atom $i$.
In reality, $F \ne 0$, and 
because $F \ge 0$, we can optimize the force-field parameters
by minimizing $F$ with respect to these parameters in the main-chain
torsion-energy term in Eq.~(\ref{ene_torsion3}).
In practice, we perform a minimization simulation in the main-chain
torsion-energy force-field 
parameter space for this
minimization \cite{SO1,SO2,SO3}.

\section{Results and Discussion}
\label{results}
\subsection{An example of the amino-acid-dependent force-field
parameter optimizations}
We present the results of our optimizations of
the force-field parameters 
$V_1(\Phi_{\rm MC}^{(k)})$ for the main-chain angles
$\Phi_{\rm MC}^{(k)}=$ 
$\psi^{(k)}$ (N-C$_{\alpha}$-C-N) and 
$\psi'^{(k)}$ (C$_{\beta}$-C$_{\alpha}$-C-N) 
in Eq.~(\ref{ene_torsion3}).
We did this for the case of AMBER ff03 force field.
We determined these
$V_1(\Phi_{\rm MC}^{(k)})$ 
values for the 19 amino-acid residues except for proline.
  
At first, we chose 100 PDB files with resolution 2.0 \AA~or better, 
with sequence similarity of amino acid 30.0 \% or lower,  
and with less than 200 residues (the average number of residues is 117.0) from PDB-REPRDB \cite{REPRDB} (see Table~\ref{PDB_list}).
We then refined these selected 100 structures.
Generally, data from X-ray experiments do not have coordinates for hydrogen atoms.
Therefore, we have to add hydrogen coordinates.
Many protein simulation software packages provide with routines that add hydrogen atoms to the PDB 
coordinates.
We used the AMBER11 program package \cite{amber_prog11}.   
We thus minimized the total potential 
energy $E_{\rm total} = E_{\rm conf} + E_{\rm solv} + E_{\rm constr}$ 
with respect to the coordinates for each proten conformation, 
where $E_{\rm constr}$ is the harmonic constraint 
energy 
term ($E_{\rm constr} = \sum_{\rm heavy~atom} K_x (\vec{x} - \vec{x}_0)^2$), 
and $E_{\rm solv}$ is the solvation energy term.
Here, $K_x$ is the force constant of the restriction 
and $\vec{x}_0$ are the original coordinate vectors 
of heavy atoms in PDB.
As one can see from $E_{\rm constr}$, the coordinates of 
hydrogen atoms will be mainly adjusted, 
but unnatural heavy-atom coordinates will also be modified.
We performed this minimization for all the 100 protein structures 
separately and obtained 
100 refined structures by using $K_x = 100$ (kcal/mol).
As for the solvation energy term $E_{\rm solv}$, we used the GB/SA solvent 
included in the AMBER program package ($igb = 5$ and $gbsa = 1$)
\cite{gbsa_igb5,gbsa_sa1}. 

For these refined protein structures, we performed the 
optimization of force-field parameters $V_1^{(k)}$ of 
$\psi$ and $\psi'$ angles for AMBER ff03 force field by 
using the fucntion $F$ in Eq.~(\ref{F}) as 
the total potential energy 
function ($E_{\rm total} = E_{\rm conf} + E_{\rm solv}$) for the Monte Carlo simulations in the
parameter space.
Here, we used AMBER11 \cite{amber_prog11} for the force calculations
in Eq.~(\ref{F1}).  
We have to optimize the 38 ($=2 \times 19$) parameters simultaneously
by the simulations in 38 parameters.  However, here,
for simplicity, we just optimized two 
parameters,  
$V_1(\psi^{(k)})$ and $V_1(\psi'^{(k)})$,  
for each amino-acid residue $k$ separately,
keeping the other $V_1$ values as the original values.
In order to obtain the optimal parameters, we performed Monte Carlo simulations 
of two parameters ($V_1$ of $\psi$ and $\psi'$) for 
the 19 amino-acid residues except for proline. 
In Table \ref{table-opt-parameters}, the optimized parameters 
are listed.

\subsection{Test simulations with two peptides}
In order to check the force-field parameters obtained by our optimization method, 
we performed the folding simulations using two peptides, namely, C-peptide of ribonuclease A and 
the C-terminal fragment of the B1 domain of streptococcal protein G, which is sometimes referred to as 
G-peptide \cite{gpep3}.
The C-peptide has 13 residues and its amino-acid sequence is
Lys-Glu$^-$-Thr-Ala-Ala-Ala-Lys$^+$-Phe-Glu-Arg$^+$-Gln-His$^+$-Met.
This peptide has been extensively studied by experiments and is known
to form an $\alpha$-helix structure \cite{buzz1,buzz2}.
Because the charges at peptide termini are known to affect
helix stability \cite{buzz1,buzz2}, 
the N and C termini of the peptide was blocked with acetyl and N-methyl groups, respectively.
The G-peptide has 16 residues and its amino-acid sequence is
Gly-Glu$^-$-Trp-Thr-Tyr-Asp$^-$-Asp$^-$-Ala-Thr-Lys$^+$-Thr-Phe-Thr-Val-Thr-Glu$^-$.
The termini were kept as the usual zwitter ionic states, following the
experimental conditions \cite{gpep3,gpep1,gpep2}.
This peptide is known to form a $\beta$-hairpin structure by experiments
\cite{gpep3,gpep1,gpep2}. 

For the folding simulations, we used replica-exchange molecular 
dynamics (REMD) \cite{REMD}.
REMD is one of the generalized-ensemble algorithms, and has high 
conformational sampling efficiency by 
allowing configurations to heat up and cool down while maintaining proper Boltzmann distributions.
We used the AMBER11 program package \cite{amber_prog11}.
The unit time step was set to 2.0 fs, and the bonds involving 
hydrogen atoms were constrained by SHAKE algorithm \cite{SHAKE}.
Each simulation was carried out for 30.0 ns (hence, it consisted of
15,000,000 MD steps) with 16 replicas by using Langevin dynamics.
The exchange procedure for each replica were performed every 3,000 MD steps.
The temperature was distributed exponentially: 
650, 612, 577, 544, 512, 483, 455, 
428, 404, 380, 358, 338, 318, 300, 282, and 266 K.
As for solvent effects, we used the GB/SA model 
in the AMBER program package ($igb = 5$ and $gbsa = 1$)
\cite{gbsa_igb5,gbsa_sa1}. 
The initial conformations for each peptide were fully
extended ones for all the replicas.
The REMD simulations were performed 
with different sets of randomly generated initial velocities
for each replica.

\begin{figure}
\begin{center}
\resizebox*{8cm}{!}{\includegraphics{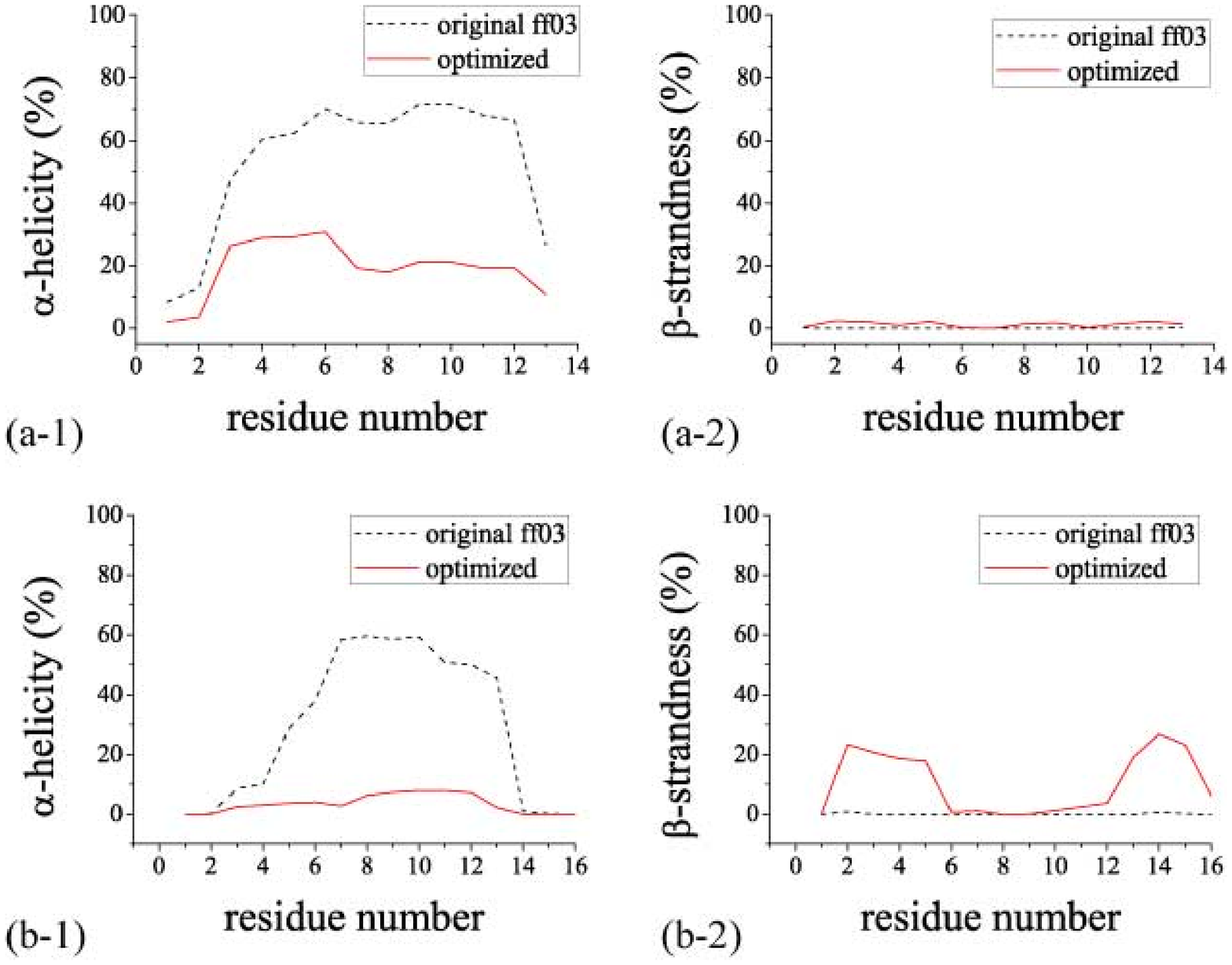}}%
\caption{$\alpha$-helicity (a-1) and $\beta$-strandness (a-2) of C-peptide 
and $\alpha$-helicity (b-1) and $\beta$-strandness (b-2) 
of G-peptide as functions of the residue number at 300 K. These values 
were obtained from the REMD simulations.
Normal and dotted curves stand for the optimized and the
original AMBER ff03 force fields, respectivery.}
\label{fig_secondary_alpha_beta}
\end{center}
\end{figure}

\begin{figure}
\begin{center}
\resizebox*{8cm}{!}{\includegraphics{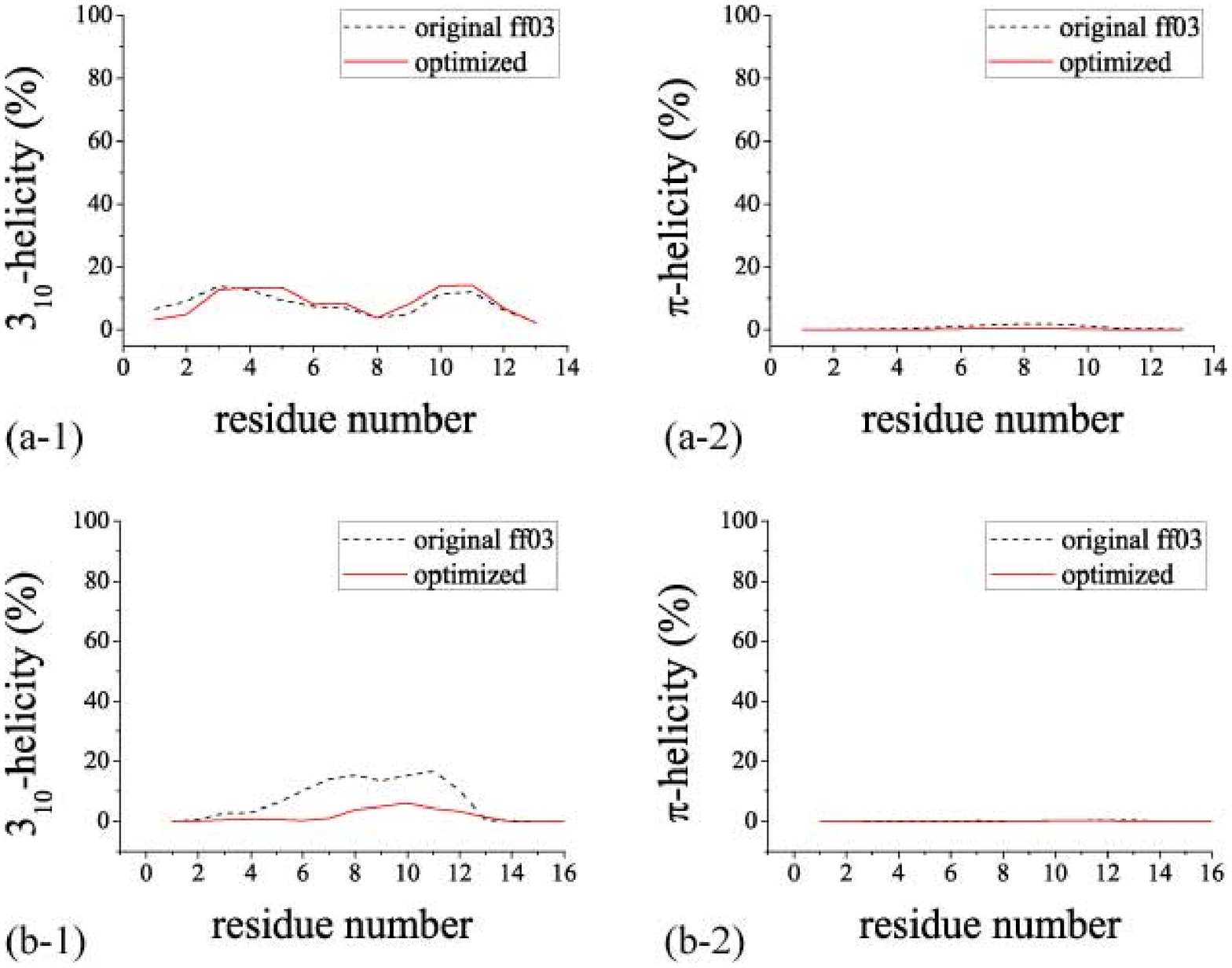}}%
\caption{3$_{10}$-helicity (a-1) and $\pi$-helicity (a-2) of C-peptide 
and 3$_{10}$-helicity (b-1) and $\pi$-helicity (b-2) 
of G-peptide as functions of the residue number at 300 K. These values 
were obtained from the REMD simulations.
Normal and dotted curves stand for the optimized and the
original AMBER ff03 force fields, respectivery.}
\label{fig_secondary_310_pi}
\end{center}
\end{figure}

In Fig.~\ref{fig_secondary_alpha_beta}, $\alpha$-helicity and 
$\beta$-strandness of two peptides obtained 
from the REMD simulations are shown.
We checked the secondary-structure formations by using the DSSP 
program \cite{DSSP},
which is based on the formations of the intra-main-chain hydrogen bonds.
As is shown in Fig.~\ref{fig_secondary_alpha_beta}, 
for the original AMBER ff03 force field, the $\alpha$-helicity is 
clearly higher than 
the $\beta$-strandness not only in C-peptide but also in G-peptide.
Namely, the original AMBER ff03 force field clearly favors 
$\alpha$-helix and does not favor 
$\beta$-structure.
On the other hand, for the optimized force field, in the case 
of C-peptide, the $\alpha$-helicity is 
higher than the $\beta$-strandness, and in the case of G-peptide, 
the $\beta$-strandness is higher than the $\alpha$-helicity.
We conclude that these results obtained from the optimized force field 
are in better agreement with 
the experimental results in comparison with the original force field.
In Fig.~\ref{fig_secondary_310_pi}, 3$_{10}$-helicity and $\pi$-helicity 
of two peptides obtained from 
the REMD simulations are shown.
For 3$_{10}$ helicity, there is no large difference for both force fields 
in C-peptide, and in the case of G-peptide, 
the value of the optimized force field slightly decreases in comparison 
with the original force field.
$\pi$-helicity has almost no value in the both cases of the original 
and optimized force fields in two peptides.

\begin{figure}
\begin{center}
\resizebox*{8cm}{!}{\includegraphics{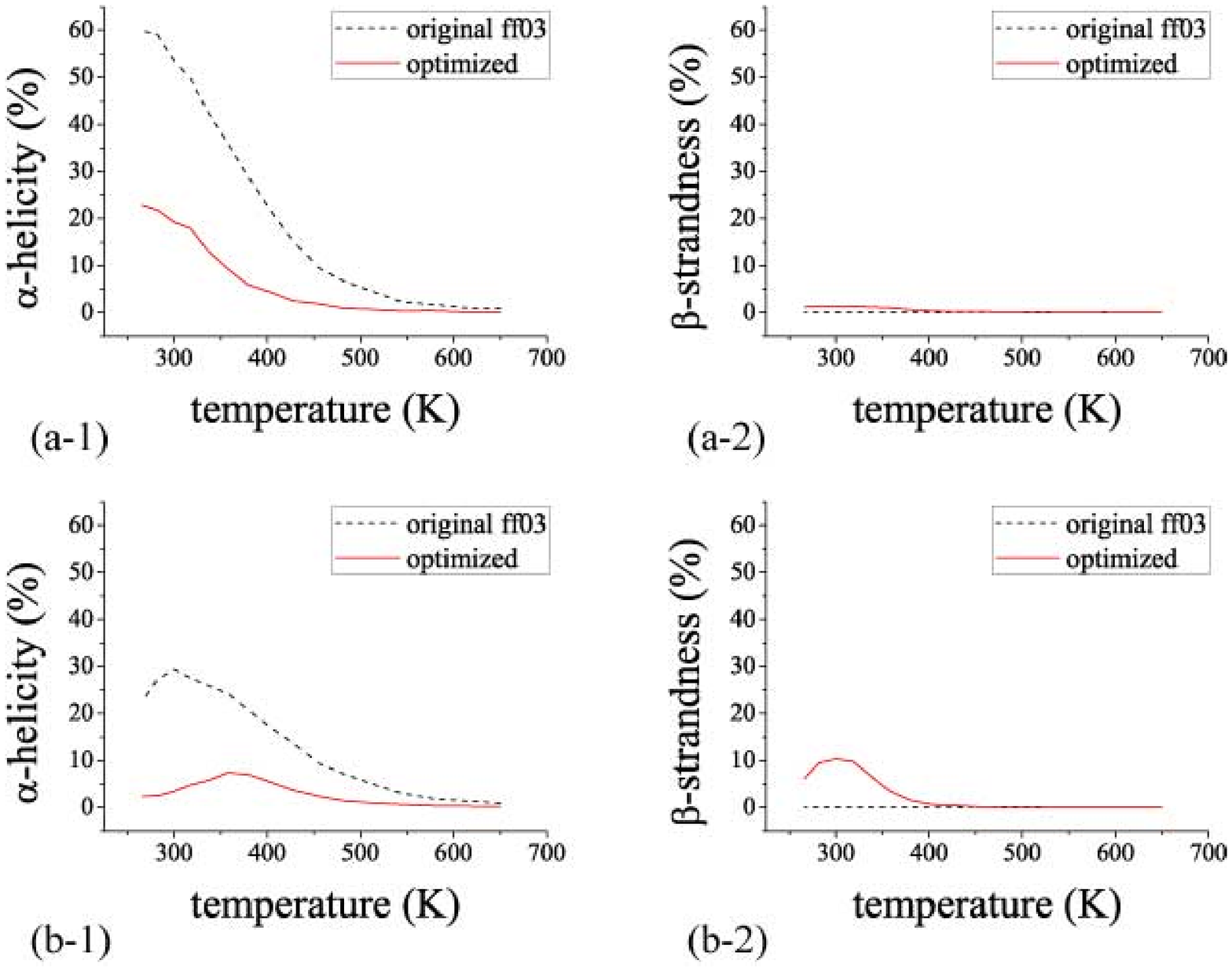}}%
\caption{$\alpha$-helicity (a-1) and $\beta$-strandness (a-2) of C-peptide 
and $\alpha$-helicity (b-1) and $\beta$-strandness (b-2) 
of G-peptide as functions of temperature. These values were
obtained from the REMD simulations.
Normal and dotted curves stand for the optimized and the 
original AMBER ff03 force fields, respectivery.}
\label{fig_temp_secondary_alpha_beta}
\end{center}
\end{figure}

\begin{figure}
\begin{center}
\resizebox*{8cm}{!}{\includegraphics{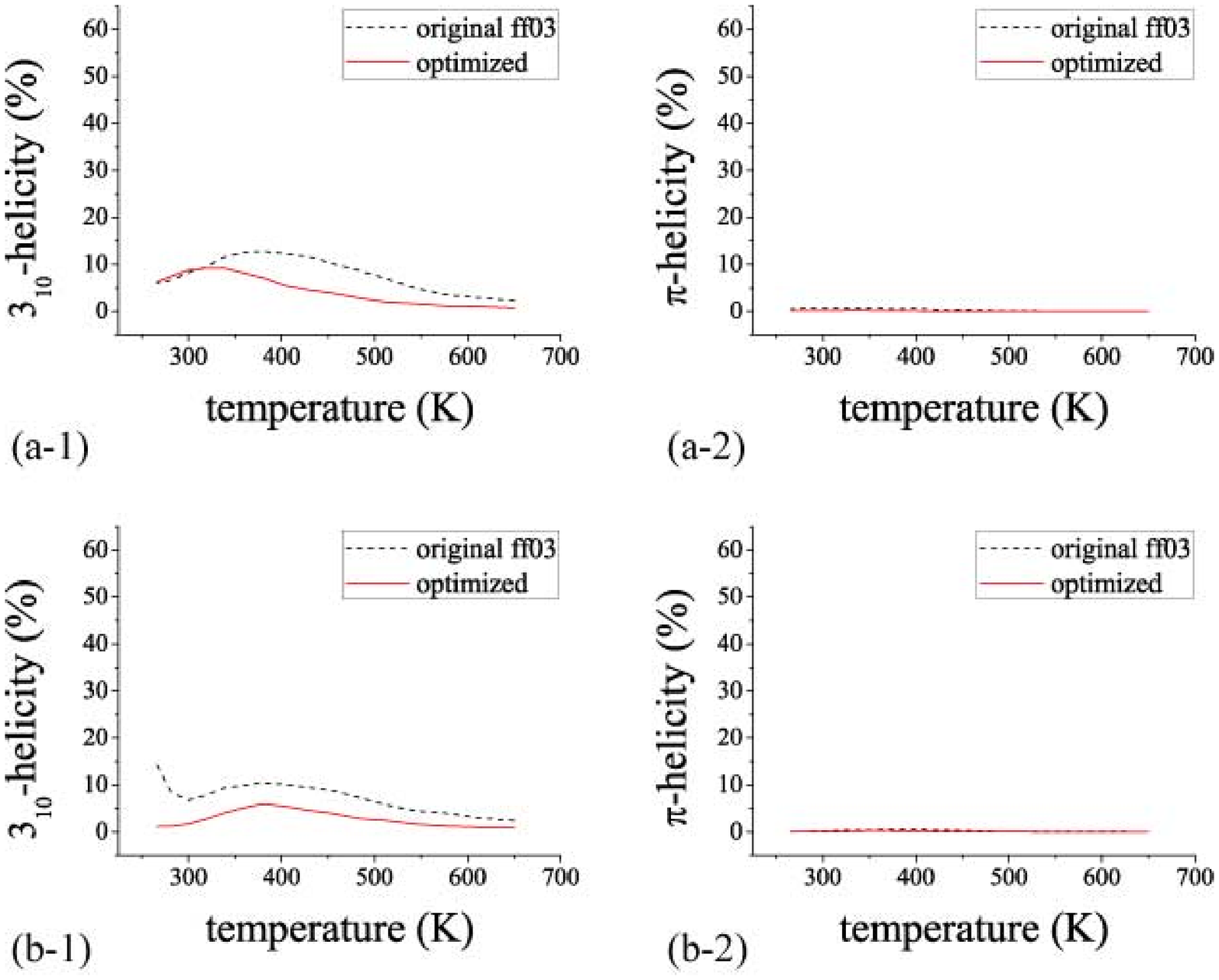}}%
\caption{3$_{10}$-helicity (a-1) and $\pi$-helicity (a-2) of C-peptide 
and 3$_{10}$-helicity (b-1) and $\pi$-helicity (b-2)
of G-peptide as functions of temperature. These values were
obtained from the REMD simulations.
Normal and dotted curves stand for the optimized and the
original AMBER ff03 force fields, respectivery.}
\label{fig_temp_secondary_310_pi}
\end{center}
\end{figure}

In Fig.~\ref{fig_temp_secondary_alpha_beta}, $\alpha$-helicity and 
$\beta$-strandness as functions of temperature for the
two peptides obtained from the REMD simulations are shown.
For $\alpha$-helicity, the values of both force fields decrease gradually from low temperature 
to high temperature in the case of C-peptide.
On the other hand, in the case of G-peptide, there are small peaks 
at around 300 K and 358 K 
for the original and optimized force fields, respectively.
For $\beta$-strandness, in the case of C-peptide, it is almost zero 
for both force fields.
In the case of G-peptide, for the optimized force field, there is clearly 
a peak around 300 K.
In Fig.~\ref{fig_temp_secondary_310_pi}, 3$_{10}$-helicity and 
$\pi$-helicity of the two peptides as functions of temperature are shown.
For 3$_{10}$-helicity, in the case of both peptides, the values of 
the optimized force field are lower 
than the original force field as a whole except around low temperature 
in C-peptide.
For $\pi$-helicity, it is almost zero for both force fields 
in the two peptides.

\begin{figure}
\begin{center}
\resizebox*{8cm}{!}{\includegraphics{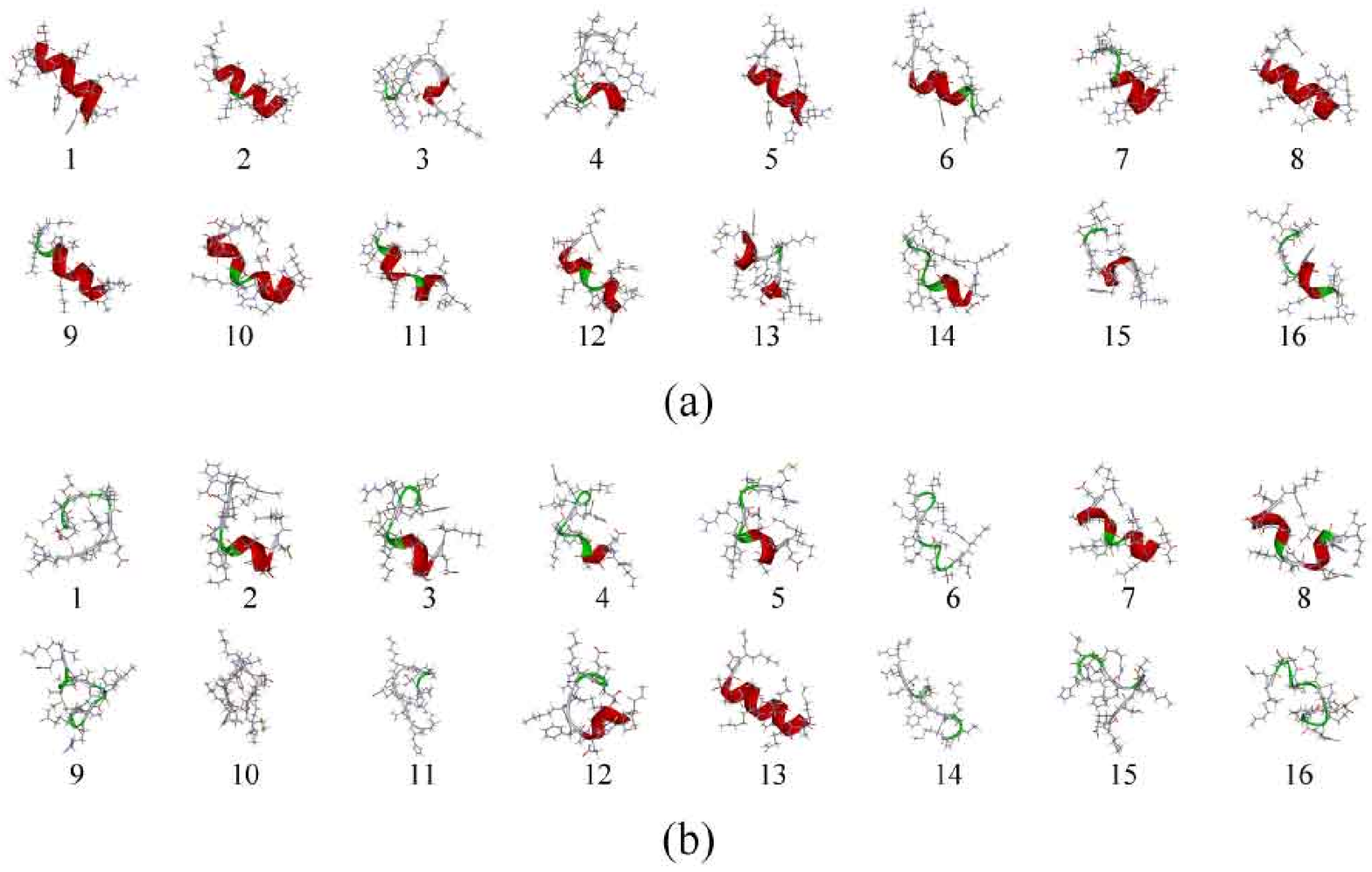}}%
\caption{Lowest-energy conformations of C-peptide obtained for
each replica from the REMD simulations. 
(a) and (b) are the results of the original AMBER ff03 and the
optimized force fields, respectively.
The conformations are ordered in the increasing order of energy.
The figures were created with DS Visualizer \cite{DSV2}.}
\label{fig_str_cp}
\end{center}
\end{figure}

\begin{figure}
\begin{center}
\resizebox*{8cm}{!}{\includegraphics{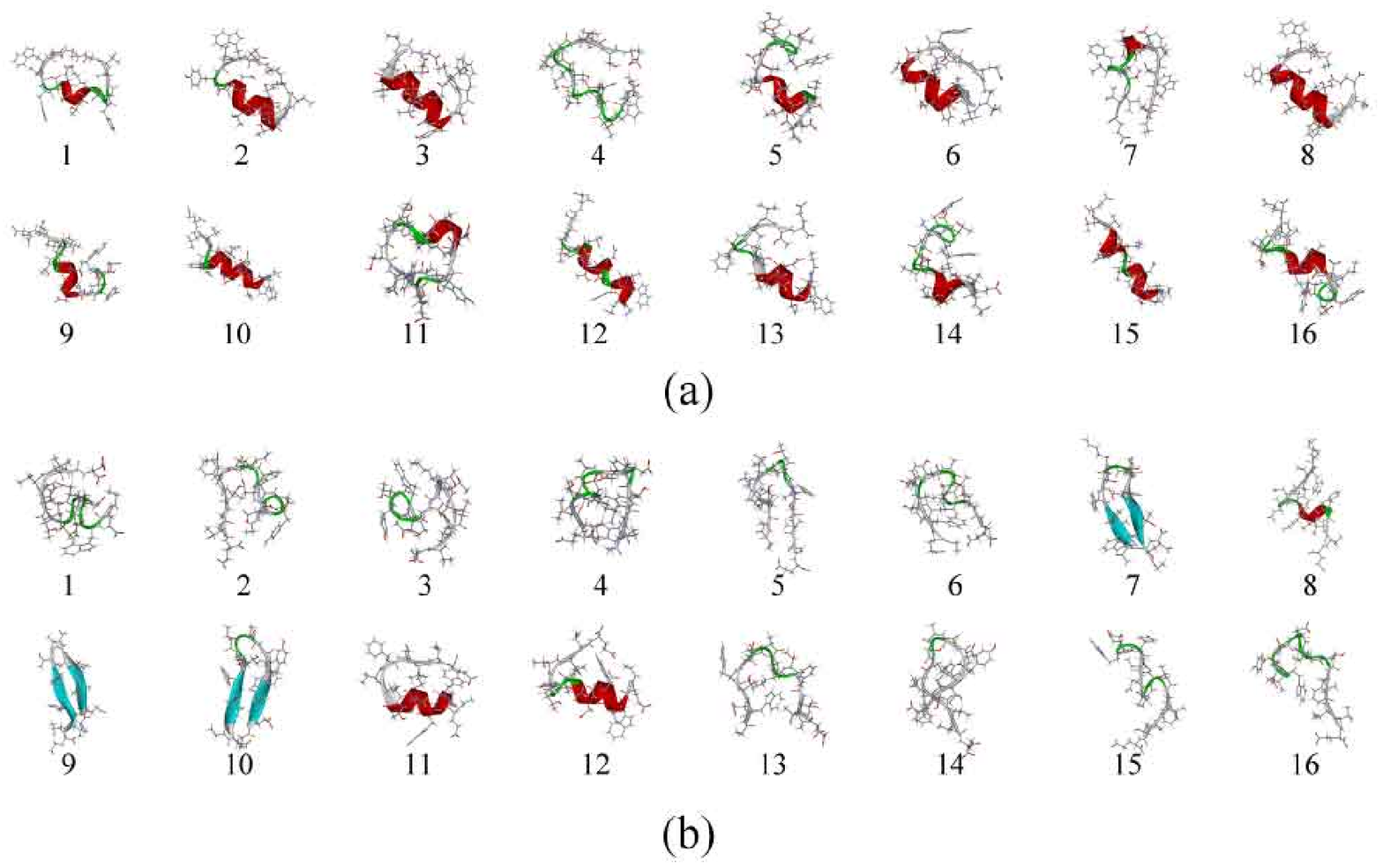}}%
\caption{Lowest-energy conformations of G-peptide obtained for
each replica from the REMD simulations. 
(a) and (b) are the results of the original AMBER ff03 and the
optimized force fields, respectively.
The conformations are ordered in the increasing order of energy.
The figures were created with DS Visualizer \cite{DSV2}.}
\label{fig_str_gp}
\end{center}
\end{figure}

In Fig.~\ref{fig_str_cp}, the lowest-energy conformations of C-peptide 
obtained from the REMD simulations 
in the case of the original and the optimized force fields are shown.
In the case of the original force field, all the conformations have helices.
No. 3 has only 3$_{10}$-helix, No. 13 has both $\alpha$-helix 
and 3$_{10}$-helix, and the rest of the conformations have only 
$\alpha$-helix.
In the case of the optimized force field, seven conformations 
(Nos. 2, 4, 5, 7, 8, 12, and 13) have helices.
Nos. 2, 5, 7, 13 have only $\alpha$-helix, Nos. 4, 12 have 
only 3$_{10}$-helix, and No. 8 has both $\alpha$-helix and 
3$_{10}$-helix.
Additionally, there is one $\beta$-bridge structure in No. 10.
In Fig.~\ref{fig_str_gp}, the lowest-energy conformations 
of G-peptide are shown.
In the case of the original force field, all the conformations 
except for No. 4 have helices.
No. 6 has both $\alpha$-helix and 3$_{10}$-helix, Nos. 5, 7, 11 
have only 3$_{10}$-helix, and the rest have only $\alpha$-helix.
In the case of the optimized force field, Nos. 11 and 12 have $\alpha$-helix, No. 8 has 3$_{10}$-helix, 
No. 5 has $\beta$-bridge, and Nos. 7, 9, 10 have $\beta$-strand.
These results clearly show that the optimized force field 
favors helix structure much less than 
the original force field, and, additionally, in the case of G-peptide, 
slightly favors $\beta$-structure.

\section{Conclusions}
The main-chain torsion-energy terms are the most problematic
terms in the force field for protein systems.  
We therefore concentrate our attention on these terms in 
order to obtain optimal protein force field.  In this
article, we proposed amino-acid-dependent main-chain
torsion-energy terms in the force field for protein systems. 
This generalization gives more freedom to the force-field
optimization problem.
In principle, we can introduce amino-acid dependence
on any force-field term.  The present work introduced
this dependence on even the main-chain torsion-energy
terms, which previously had been treated independent
of the amino-acid residue type.

As an example of the present general formalism,
we modified the AMBER ff03 force field so that
the $V_1$ parameters of the main-chain $\psi$
and $\psi^{\prime}$ angles may be amino-acid
dependent except for proline (hence, 38 parameters
were optimized).  Although preliminary because we
did not optimize the 38 parameters simulataneously,
our optimized parameters already gave structures
more consistent with the experimental implications
than the original AMBER force field in the folding simulations
of two small peptides.

We can easily apply the present formulations to other
popular force fields such as AMBER ff99SB,
CHARMM22/CMAP, etc.  This will be our future work.

\begin{acknowledgments}
This article is dedicated to the 90-th birthday of Professor Harold
A. Scheraga.
The computations were performed on the computers at the Research Center for Computational Science, 
Institute for Molecular Science, Information Technology Center, Nagoya University, and Center for 
Computational Sciences, University of Tsukuba.
This work was supported, in part, by 
the Grants-in-Aid for the Academic Frontier Project, ``Intelligent Information Science'', 
for Scientific Research on Innovative Areas (``Fluctuations and Biological Functions'' ),
and for the Next Generation Super Computing Project, Nanoscience Program
and Computational Materials Science Initiative 
from the Ministry of Education, Culture, Sports, Science 
and Technology (MEXT), Japan.
\end{acknowledgments}

~~\\
~~\\

\noindent
{\bf REFERENCES}


\begin{thebibliography}{49}%
\makeatletter
\providecommand \@ifxundefined [1]{%
 \@ifx{#1\undefined}
}%
\providecommand \@ifnum [1]{%
 \ifnum #1\expandafter \@firstoftwo
 \else \expandafter \@secondoftwo
 \fi
}%
\providecommand \@ifx [1]{%
 \ifx #1\expandafter \@firstoftwo
 \else \expandafter \@secondoftwo
 \fi
}%
\providecommand \natexlab [1]{#1}%
\providecommand \enquote  [1]{``#1''}%
\providecommand \bibnamefont  [1]{#1}%
\providecommand \bibfnamefont [1]{#1}%
\providecommand \citenamefont [1]{#1}%
\providecommand \href@noop [0]{\@secondoftwo}%
\providecommand \href [0]{\begingroup \@sanitize@url \@href}%
\providecommand \@href[1]{\@@startlink{#1}\@@href}%
\providecommand \@@href[1]{\endgroup#1\@@endlink}%
\providecommand \@sanitize@url [0]{\catcode `\\12\catcode `\$12\catcode
  `\&12\catcode `\#12\catcode `\^12\catcode `\_12\catcode `\%12\relax}%
\providecommand \@@startlink[1]{}%
\providecommand \@@endlink[0]{}%
\providecommand \url  [0]{\begingroup\@sanitize@url \@url }%
\providecommand \@url [1]{\endgroup\@href {#1}{\urlprefix }}%
\providecommand \urlprefix  [0]{URL }%
\providecommand \Eprint [0]{\href }%
\providecommand \doibase [0]{http://dx.doi.org/}%
\providecommand \selectlanguage [0]{\@gobble}%
\providecommand \bibinfo  [0]{\@secondoftwo}%
\providecommand \bibfield  [0]{\@secondoftwo}%
\providecommand \translation [1]{[#1]}%
\providecommand \BibitemOpen [0]{}%
\providecommand \bibitemStop [0]{}%
\providecommand \bibitemNoStop [0]{.\EOS\space}%
\providecommand \EOS [0]{\spacefactor3000\relax}%
\providecommand \BibitemShut  [1]{\csname bibitem#1\endcsname}%
\let\auto@bib@innerbib\@empty
\bibitem [{\citenamefont {Liwo}\ \emph {et~al.}(2008)\citenamefont {Liwo},
  \citenamefont {Czaplewski}, \citenamefont {Stanislaw},\ and\ \citenamefont
  {Scheraga}}]{ScheragaRev1}%
  \BibitemOpen
  \bibfield  {author} {\bibinfo {author} {\bibfnamefont {A.}~\bibnamefont
  {Liwo}}, \bibinfo {author} {\bibfnamefont {C.}~\bibnamefont {Czaplewski}},
  \bibinfo {author} {\bibfnamefont {O.}~\bibnamefont {Stanislaw}}, \ and\
  \bibinfo {author} {\bibfnamefont {H.~A.}\ \bibnamefont {Scheraga}},\
  }\bibfield  {title} {\enquote {\bibinfo {title} {Computational techniques for
  efficient conformational sampling of proteins},}\ }\href@noop {} {\bibfield
  {journal} {\bibinfo  {journal} {Curr. Opin. Struct. Biol.}\ }\textbf
  {\bibinfo {volume} {18}},\ \bibinfo {pages} {134--139} (\bibinfo {year}
  {2008})}
\bibitem [{\citenamefont {Scheraga}(2011)}]{Scheragarev2}%
  \BibitemOpen
  \bibfield  {author} {\bibinfo {author} {\bibfnamefont {H.~A.}\ \bibnamefont
  {Scheraga}},\ }\bibfield  {title} {\enquote {\bibinfo {title} {Respice,
  adspice, and prospice},}\ }\href@noop {} {\bibfield  {journal} {\bibinfo
  {journal} {Ann. Rev. Biophys.}\ }\textbf {\bibinfo {volume} {40}},\ \bibinfo
  {pages} {1--39} (\bibinfo {year} {2011})}
\bibitem [{\citenamefont {Hansmann}\ and\ \citenamefont
  {Okamoto}(1999)}]{HOrev}%
  \BibitemOpen
  \bibfield  {author} {\bibinfo {author} {\bibfnamefont {U.~H.~E.}\
  \bibnamefont {Hansmann}}\ and\ \bibinfo {author} {\bibfnamefont
  {Y.}~\bibnamefont {Okamoto}},\ }\bibfield  {title} {\enquote {\bibinfo
  {title} {New monte carlo algorithms for protein folding},}\ }\href@noop {}
  {\bibfield  {journal} {\bibinfo  {journal} {Curr. Opin. Struct. Biol.}\
  }\textbf {\bibinfo {volume} {9}},\ \bibinfo {pages} {177--183} (\bibinfo
  {year} {1999})}
\bibitem [{\citenamefont {Mitsutake}, \citenamefont {Sugita},\ and\
  \citenamefont {Okamoto}(2001)}]{GEA}%
  \BibitemOpen
  \bibfield  {author} {\bibinfo {author} {\bibfnamefont {A.}~\bibnamefont
  {Mitsutake}}, \bibinfo {author} {\bibfnamefont {Y.}~\bibnamefont {Sugita}}, \
  and\ \bibinfo {author} {\bibfnamefont {Y.}~\bibnamefont {Okamoto}},\
  }\bibfield  {title} {\enquote {\bibinfo {title} {Generalized-ensemble
  algorithms for molecular simulations of biopolymers},}\ }\href@noop {}
  {\bibfield  {journal} {\bibinfo  {journal} {Biopolymers}\ }\textbf {\bibinfo
  {volume} {60}},\ \bibinfo {pages} {96--123} (\bibinfo {year} {2001})}
\bibitem [{\citenamefont {Cornell}\ \emph {et~al.}(1995)\citenamefont
  {Cornell}, \citenamefont {Cieplak}, \citenamefont {Bayly}, \citenamefont
  {Gould}, \citenamefont {Kenneth M.~Merz}, \citenamefont {Ferguson},
  \citenamefont {Spellmeyer}, \citenamefont {Fox}, \citenamefont {Caldwell},\
  and\ \citenamefont {Kollman}}]{parm94}%
  \BibitemOpen
  \bibfield  {author} {\bibinfo {author} {\bibfnamefont {W.~D.}\ \bibnamefont
  {Cornell}}, \bibinfo {author} {\bibfnamefont {P.}~\bibnamefont {Cieplak}},
  \bibinfo {author} {\bibfnamefont {C.~I.}\ \bibnamefont {Bayly}}, \bibinfo
  {author} {\bibfnamefont {I.~R.}\ \bibnamefont {Gould}}, \bibinfo {author}
  {\bibfnamefont {J.}~\bibnamefont {Kenneth M.~Merz}}, \bibinfo {author}
  {\bibfnamefont {D.~M.}\ \bibnamefont {Ferguson}}, \bibinfo {author}
  {\bibfnamefont {D.~C.}\ \bibnamefont {Spellmeyer}}, \bibinfo {author}
  {\bibfnamefont {T.}~\bibnamefont {Fox}}, \bibinfo {author} {\bibfnamefont
  {J.~W.}\ \bibnamefont {Caldwell}}, \ and\ \bibinfo {author} {\bibfnamefont
  {P.~A.}\ \bibnamefont {Kollman}},\ }\bibfield  {title} {\enquote {\bibinfo
  {title} {A second generation force field for the simulation of proteins,
  nucleic acids, and organic molecules},}\ }\href@noop {} {\bibfield  {journal}
  {\bibinfo  {journal} {J. Am. Chem. Soc.}\ }\textbf {\bibinfo {volume}
  {117}},\ \bibinfo {pages} {5179--5197} (\bibinfo {year} {1995})}
\bibitem [{\citenamefont {Kollman}\ \emph
  {et~al.}(1997{\natexlab{a}})\citenamefont {Kollman}, \citenamefont {Dixon},
  \citenamefont {Cornell}, \citenamefont {Fox}, \citenamefont {Chipot},\ and\
  \citenamefont {Pohorille}}]{parm96_2}%
  \BibitemOpen
  \bibfield  {author} {\bibinfo {author} {\bibfnamefont {P.~A.}\ \bibnamefont
  {Kollman}}, \bibinfo {author} {\bibfnamefont {R.}~\bibnamefont {Dixon}},
  \bibinfo {author} {\bibfnamefont {W.}~\bibnamefont {Cornell}}, \bibinfo
  {author} {\bibfnamefont {T.}~\bibnamefont {Fox}}, \bibinfo {author}
  {\bibfnamefont {C.}~\bibnamefont {Chipot}}, \ and\ \bibinfo {author}
  {\bibfnamefont {A.}~\bibnamefont {Pohorille}},\ }\href@noop {} {\emph
  {\bibinfo {title} {Computer simulations of biological systems}}},\ edited by\
  \bibinfo {editor} {\bibfnamefont {W.~F.}\ \bibnamefont {van Gunsteren}}\ and\
  \bibinfo {editor} {\bibfnamefont {P.~K.}\ \bibnamefont {Weiner}},\
  Vol.~\bibinfo {volume} {3}\ (\bibinfo  {publisher} {ESCOM},\ \bibinfo
  {address} {Dordrecht},\ \bibinfo {year} {1997})\ pp.\ \bibinfo {pages}
  {83--96}
\bibitem [{\citenamefont {Wang}, \citenamefont {Cieplak},\ and\ \citenamefont
  {Kollman}(2000)}]{parm99}%
  \BibitemOpen
  \bibfield  {author} {\bibinfo {author} {\bibfnamefont {J.}~\bibnamefont
  {Wang}}, \bibinfo {author} {\bibfnamefont {P.}~\bibnamefont {Cieplak}}, \
  and\ \bibinfo {author} {\bibfnamefont {P.~A.}\ \bibnamefont {Kollman}},\
  }\bibfield  {title} {\enquote {\bibinfo {title} {How well does a restrained
  electrostatic potential (resp) model perform in calculating conformational
  energies of organic and biological molecules?}}\ }\href@noop {} {\bibfield
  {journal} {\bibinfo  {journal} {J. Comput. Chem.}\ }\textbf {\bibinfo
  {volume} {21}},\ \bibinfo {pages} {1049--1074} (\bibinfo {year} {2000})}
\bibitem [{\citenamefont {Hornak}\ \emph {et~al.}(2006)\citenamefont {Hornak},
  \citenamefont {Abel}, \citenamefont {Strockbine}, \citenamefont {Roitberg},\
  and\ \citenamefont {Simmerling}}]{parm99SB}%
  \BibitemOpen
  \bibfield  {author} {\bibinfo {author} {\bibfnamefont {V.}~\bibnamefont
  {Hornak}}, \bibinfo {author} {\bibfnamefont {A.}~\bibnamefont {Abel},
  \bibfnamefont {R.~Okur}}, \bibinfo {author} {\bibfnamefont {B.}~\bibnamefont
  {Strockbine}}, \bibinfo {author} {\bibfnamefont {A.}~\bibnamefont
  {Roitberg}}, \ and\ \bibinfo {author} {\bibfnamefont {C.}~\bibnamefont
  {Simmerling}},\ }\bibfield  {title} {\enquote {\bibinfo {title} {Comparison
  of multiple amber force fields and development of improved protein backbone
  parameters},}\ }\href@noop {} {\bibfield  {journal} {\bibinfo  {journal}
  {Proteins}\ }\textbf {\bibinfo {volume} {65}},\ \bibinfo {pages} {712--725}
  (\bibinfo {year} {2006})}
\bibitem [{\citenamefont {Duan}\ \emph
  {et~al.}(2003{\natexlab{a}})\citenamefont {Duan}, \citenamefont {Wu},
  \citenamefont {Chowdhury}, \citenamefont {Lee}, \citenamefont {Xiong},
  \citenamefont {Zhang}, \citenamefont {Yang}, \citenamefont {Cieplak},
  \citenamefont {Luo},\ and\ \citenamefont {Lee}}]{parm03}%
  \BibitemOpen
  \bibfield  {author} {\bibinfo {author} {\bibfnamefont {Y.}~\bibnamefont
  {Duan}}, \bibinfo {author} {\bibfnamefont {C.}~\bibnamefont {Wu}}, \bibinfo
  {author} {\bibfnamefont {S.}~\bibnamefont {Chowdhury}}, \bibinfo {author}
  {\bibfnamefont {M.~C.}\ \bibnamefont {Lee}}, \bibinfo {author} {\bibfnamefont
  {G.}~\bibnamefont {Xiong}}, \bibinfo {author} {\bibfnamefont
  {W.}~\bibnamefont {Zhang}}, \bibinfo {author} {\bibfnamefont
  {R.}~\bibnamefont {Yang}}, \bibinfo {author} {\bibfnamefont {P.}~\bibnamefont
  {Cieplak}}, \bibinfo {author} {\bibfnamefont {R.}~\bibnamefont {Luo}}, \ and\
  \bibinfo {author} {\bibfnamefont {T.}~\bibnamefont {Lee}},\ }\bibfield
  {title} {\enquote {\bibinfo {title} {A point-charge force field for molecular
  mechanics simulations of proteins based on condensed-phase quantum mechanical
  calculations},}\ }\href@noop {} {\bibfield  {journal} {\bibinfo  {journal}
  {J. Comput. Chem.}\ }\textbf {\bibinfo {volume} {24}},\ \bibinfo {pages}
  {1999--2012} (\bibinfo {year} {2003}{\natexlab{a}})}
\bibitem [{\citenamefont {MacKerell~Jr}\ \emph {et~al.}(1998)\citenamefont
  {MacKerell~Jr}, \citenamefont {Bashford}, \citenamefont {Bellott},
  \citenamefont {Dunbrack}, \citenamefont {Evanseck}, \citenamefont {Field},
  \citenamefont {Fischer}, \citenamefont {Gao}, \citenamefont {Guo},
  \citenamefont {Ha}, \citenamefont {Joseph-McCarthy}, \citenamefont {Kuchnir},
  \citenamefont {Kuczera}, \citenamefont {Lau}, \citenamefont {Mattos},
  \citenamefont {Michnick}, \citenamefont {Ngo}, \citenamefont {Nguyen},
  \citenamefont {Prodhom}, \citenamefont {Reiher}, \citenamefont {Roux},
  \citenamefont {Schlenkrich}, \citenamefont {Smith}, \citenamefont {Stote},
  \citenamefont {Watanabe}, \citenamefont {Wiorkiewicz-Kuczera}, \citenamefont
  {Yin},\ and\ \citenamefont {Karplus}}]{charmm}%
  \BibitemOpen
  \bibfield  {author} {\bibinfo {author} {\bibfnamefont {A.~D.}\ \bibnamefont
  {MacKerell~Jr}}, \bibinfo {author} {\bibfnamefont {D.}~\bibnamefont
  {Bashford}}, \bibinfo {author} {\bibfnamefont {M.}~\bibnamefont {Bellott}},
  \bibinfo {author} {\bibfnamefont {J.}~\bibnamefont {Dunbrack}, \bibfnamefont
  {R.~L.}}, \bibinfo {author} {\bibfnamefont {J.~D.}\ \bibnamefont {Evanseck}},
  \bibinfo {author} {\bibfnamefont {M.~J.}\ \bibnamefont {Field}}, \bibinfo
  {author} {\bibfnamefont {S.}~\bibnamefont {Fischer}}, \bibinfo {author}
  {\bibfnamefont {J.}~\bibnamefont {Gao}}, \bibinfo {author} {\bibfnamefont
  {H.}~\bibnamefont {Guo}}, \bibinfo {author} {\bibfnamefont {S.}~\bibnamefont
  {Ha}}, \bibinfo {author} {\bibfnamefont {D.}~\bibnamefont {Joseph-McCarthy}},
  \bibinfo {author} {\bibfnamefont {L.}~\bibnamefont {Kuchnir}}, \bibinfo
  {author} {\bibfnamefont {K.}~\bibnamefont {Kuczera}}, \bibinfo {author}
  {\bibfnamefont {F.~T.~K.}\ \bibnamefont {Lau}}, \bibinfo {author}
  {\bibfnamefont {C.}~\bibnamefont {Mattos}}, \bibinfo {author} {\bibfnamefont
  {S.}~\bibnamefont {Michnick}}, \bibinfo {author} {\bibfnamefont
  {T.}~\bibnamefont {Ngo}}, \bibinfo {author} {\bibfnamefont {D.~T.}\
  \bibnamefont {Nguyen}}, \bibinfo {author} {\bibfnamefont {B.}~\bibnamefont
  {Prodhom}}, \bibinfo {author} {\bibfnamefont {I.}~\bibnamefont {Reiher},
  \bibfnamefont {W.~E.}}, \bibinfo {author} {\bibfnamefont {B.}~\bibnamefont
  {Roux}}, \bibinfo {author} {\bibfnamefont {M.}~\bibnamefont {Schlenkrich}},
  \bibinfo {author} {\bibfnamefont {J.~C.}\ \bibnamefont {Smith}}, \bibinfo
  {author} {\bibfnamefont {J.}~\bibnamefont {Stote}, \bibfnamefont
  {R.;~Straub}}, \bibinfo {author} {\bibfnamefont {M.}~\bibnamefont
  {Watanabe}}, \bibinfo {author} {\bibfnamefont {J.}~\bibnamefont
  {Wiorkiewicz-Kuczera}}, \bibinfo {author} {\bibfnamefont {D.}~\bibnamefont
  {Yin}}, \ and\ \bibinfo {author} {\bibfnamefont {M.}~\bibnamefont
  {Karplus}},\ }\bibfield  {title} {\enquote {\bibinfo {title} {All-atom
  empirical potential for molecular modeling and dynamics studies of
  proteins},}\ }\href@noop {} {\bibfield  {journal} {\bibinfo  {journal} {J
  Phys Chem B}\ }\textbf {\bibinfo {volume} {102}},\ \bibinfo {pages}
  {3586--3616} (\bibinfo {year} {1998})}
\bibitem [{\citenamefont {MacKerell~Jr}, \citenamefont {Feig},\ and\
  \citenamefont {Brooks~III}(2004)}]{CMAP}%
  \BibitemOpen
  \bibfield  {author} {\bibinfo {author} {\bibfnamefont {A.~D.}\ \bibnamefont
  {MacKerell~Jr}}, \bibinfo {author} {\bibfnamefont {M.}~\bibnamefont {Feig}},
  \ and\ \bibinfo {author} {\bibfnamefont {C.}~\bibnamefont {Brooks~III}},\
  }\bibfield  {title} {\enquote {\bibinfo {title} {Extending the treatment of
  backbone energetics in protein force fields: limitations of gas-phase quantum
  mechanics in reproducing protein conformational distributions in molecular
  dynamics simulations},}\ }\href@noop {} {\bibfield  {journal} {\bibinfo
  {journal} {J. Comput. Chem.}\ }\textbf {\bibinfo {volume} {25}},\ \bibinfo
  {pages} {1400--1415} (\bibinfo {year} {2004})}
\bibitem [{\citenamefont {Jorgensen}, \citenamefont {Maxwell},\ and\
  \citenamefont {Tirado-Rives}(1996)}]{opls1}%
  \BibitemOpen
  \bibfield  {author} {\bibinfo {author} {\bibfnamefont {W.~L.}\ \bibnamefont
  {Jorgensen}}, \bibinfo {author} {\bibfnamefont {D.~S.}\ \bibnamefont
  {Maxwell}}, \ and\ \bibinfo {author} {\bibfnamefont {J.}~\bibnamefont
  {Tirado-Rives}},\ }\bibfield  {title} {\enquote {\bibinfo {title}
  {Development and testing of the opls all-atom force field on conformational
  energetics and properties of organic liquids},}\ }\href@noop {} {\bibfield
  {journal} {\bibinfo  {journal} {J. Am. Chem. Soc.}\ }\textbf {\bibinfo
  {volume} {118}},\ \bibinfo {pages} {11225--11236} (\bibinfo {year} {1996})}
\bibitem [{\citenamefont {Kaminski}\ \emph {et~al.}(2001)\citenamefont
  {Kaminski}, \citenamefont {Friesner}, \citenamefont {Tirado-Rives},\ and\
  \citenamefont {Jorgensen}}]{opls2}%
  \BibitemOpen
  \bibfield  {author} {\bibinfo {author} {\bibfnamefont {G.~A.}\ \bibnamefont
  {Kaminski}}, \bibinfo {author} {\bibfnamefont {R.~A.}\ \bibnamefont
  {Friesner}}, \bibinfo {author} {\bibfnamefont {J.}~\bibnamefont
  {Tirado-Rives}}, \ and\ \bibinfo {author} {\bibfnamefont {W.~L.}\
  \bibnamefont {Jorgensen}},\ }\bibfield  {title} {\enquote {\bibinfo {title}
  {Evaluation and reparametrization of the opls-aa force field for proteins via
  comparison with accurate quantum chemical calculations on peptides},}\
  }\href@noop {} {\bibfield  {journal} {\bibinfo  {journal} {J. Phys. Chem. B}\
  }\textbf {\bibinfo {volume} {105}},\ \bibinfo {pages} {6474--6487} (\bibinfo
  {year} {2001})}
\bibitem [{\citenamefont {van Gunsteren}\ \emph {et~al.}(1996)\citenamefont
  {van Gunsteren}, \citenamefont {Billeter}, \citenamefont {Eising},
  \citenamefont {H{\"u}nenberger}, \citenamefont {Kr{\"u}ger}, \citenamefont
  {Mark}, \citenamefont {Scott},\ and\ \citenamefont {Tironi}}]{gromos_v2}%
  \BibitemOpen
  \bibfield  {author} {\bibinfo {author} {\bibfnamefont {W.~F.}\ \bibnamefont
  {van Gunsteren}}, \bibinfo {author} {\bibfnamefont {S.~R.}\ \bibnamefont
  {Billeter}}, \bibinfo {author} {\bibfnamefont {A.~A.}\ \bibnamefont
  {Eising}}, \bibinfo {author} {\bibfnamefont {P.~H.}\ \bibnamefont
  {H{\"u}nenberger}}, \bibinfo {author} {\bibfnamefont {P.}~\bibnamefont
  {Kr{\"u}ger}}, \bibinfo {author} {\bibfnamefont {A.~E.}\ \bibnamefont
  {Mark}}, \bibinfo {author} {\bibfnamefont {W.~R.~P.}\ \bibnamefont {Scott}},
  \ and\ \bibinfo {author} {\bibfnamefont {I.~G.}\ \bibnamefont {Tironi}},\
  }\href@noop {} {\emph {\bibinfo {title} {Biomolecular Simulation: The
  GROMOS96 Manual and User Guide}}}\ (\bibinfo  {publisher} {Vdf
  Hochschulverlag AG an der ETH Z{\"u}rich},\ \bibinfo {address} {Z{\"u}rich},\
  \bibinfo {year} {1996})
\bibitem [{\citenamefont {Oostenbrink}\ \emph {et~al.}(2004)\citenamefont
  {Oostenbrink}, \citenamefont {Villa}, \citenamefont {Mark},\ and\
  \citenamefont {van Gunsteren}}]{gromos_v3}%
  \BibitemOpen
  \bibfield  {author} {\bibinfo {author} {\bibfnamefont {C.}~\bibnamefont
  {Oostenbrink}}, \bibinfo {author} {\bibfnamefont {A.}~\bibnamefont {Villa}},
  \bibinfo {author} {\bibfnamefont {A.~E.}\ \bibnamefont {Mark}}, \ and\
  \bibinfo {author} {\bibfnamefont {W.~F.}\ \bibnamefont {van Gunsteren}},\
  }\bibfield  {title} {\enquote {\bibinfo {title} {A biomolecular force field
  based on the free enthalpy of hydration and solvation: the gromos force-field
  parameter sets 53a5 and 53a6},}\ }\href@noop {} {\bibfield  {journal}
  {\bibinfo  {journal} {J. Comput. Chem.}\ }\textbf {\bibinfo {volume} {25}},\
  \bibinfo {pages} {1656--1676} (\bibinfo {year} {2004})}
\bibitem [{\citenamefont {Berendsen}, \citenamefont {van~der Spoel},\ and\
  \citenamefont {van Drunen}(1995)}]{gromacs1}%
  \BibitemOpen
  \bibfield  {author} {\bibinfo {author} {\bibfnamefont {H.~J.~C.}\
  \bibnamefont {Berendsen}}, \bibinfo {author} {\bibfnamefont {D.}~\bibnamefont
  {van~der Spoel}}, \ and\ \bibinfo {author} {\bibfnamefont {R.}~\bibnamefont
  {van Drunen}},\ }\bibfield  {title} {\enquote {\bibinfo {title} {Gromacs: a
  message-passing parallel molecular dynamics implementation},}\ }\href@noop {}
  {\bibfield  {journal} {\bibinfo  {journal} {Comput. Phys. Commun.}\ }\textbf
  {\bibinfo {volume} {91}},\ \bibinfo {pages} {43--56} (\bibinfo {year}
  {1995})}
\bibitem [{\citenamefont {Lindahl}, \citenamefont {Hess},\ and\ \citenamefont
  {van~der Spoel}()}]{gromacs2}%
  \BibitemOpen
  \bibfield  {author} {\bibinfo {author} {\bibfnamefont {E.}~\bibnamefont
  {Lindahl}}, \bibinfo {author} {\bibfnamefont {B.}~\bibnamefont {Hess}}, \
  and\ \bibinfo {author} {\bibfnamefont {D.}~\bibnamefont {van~der Spoel}},\
  }\bibfield  {title} {\enquote {\bibinfo {title} {Gromacs 3.0: a package for
  molecular simulation and trajectory anaylysis},}\ }\href@noop {} {\ }
\bibitem [{\citenamefont {N\'emethy}\ \emph {et~al.}(1992)\citenamefont
  {N\'emethy}, \citenamefont {Gibson}, \citenamefont {Palmer}, \citenamefont
  {Yoon}, \citenamefont {Paterlini}, \citenamefont {Zagari}, \citenamefont
  {Rumsey},\ and\ \citenamefont {Scheraga}}]{ECEPP}%
  \BibitemOpen
  \bibfield  {author} {\bibinfo {author} {\bibfnamefont {G.}~\bibnamefont
  {N\'emethy}}, \bibinfo {author} {\bibfnamefont {K.~D.}\ \bibnamefont
  {Gibson}}, \bibinfo {author} {\bibfnamefont {K.~A.}\ \bibnamefont {Palmer}},
  \bibinfo {author} {\bibfnamefont {C.~N.}\ \bibnamefont {Yoon}}, \bibinfo
  {author} {\bibfnamefont {G.}~\bibnamefont {Paterlini}}, \bibinfo {author}
  {\bibfnamefont {A.}~\bibnamefont {Zagari}}, \bibinfo {author} {\bibfnamefont
  {S.}~\bibnamefont {Rumsey}}, \ and\ \bibinfo {author} {\bibfnamefont {H.~A.}\
  \bibnamefont {Scheraga}},\ }\bibfield  {title} {\enquote {\bibinfo {title}
  {Energy parameters in polypeptides. 10. improved geometrical parameters and
  nonbonded interactions for use in the ecepp/3 algorithm, with application to
  proline-containing peptides},}\ }\href@noop {} {\bibfield  {journal}
  {\bibinfo  {journal} {J. Phys. Chem.}\ }\textbf {\bibinfo {volume} {96}},\
  \bibinfo {pages} {6472--6484} (\bibinfo {year} {1992})}
\bibitem [{\citenamefont {Arnautova}, \citenamefont {Jagielska},\ and\
  \citenamefont {Scheraga}(2006)}]{ECEPP05}%
  \BibitemOpen
  \bibfield  {author} {\bibinfo {author} {\bibfnamefont {Y.~A.}\ \bibnamefont
  {Arnautova}}, \bibinfo {author} {\bibfnamefont {A.}~\bibnamefont
  {Jagielska}}, \ and\ \bibinfo {author} {\bibfnamefont {H.~A.}\ \bibnamefont
  {Scheraga}},\ }\bibfield  {title} {\enquote {\bibinfo {title} {A new force
  field (ecepp-05) for peptides, proteins, and organic molecules},}\
  }\href@noop {} {\bibfield  {journal} {\bibinfo  {journal} {J. Phys. Chem. B}\
  }\textbf {\bibinfo {volume} {110}},\ \bibinfo {pages} {5025--5044} (\bibinfo
  {year} {2006})}
\bibitem [{\citenamefont {Kollman}\ \emph
  {et~al.}(1997{\natexlab{b}})\citenamefont {Kollman}, \citenamefont {Dixon},
  \citenamefont {Cornell}, \citenamefont {Fox}, \citenamefont {Chipot},\ and\
  \citenamefont {Pohorille}}]{parm96}%
  \BibitemOpen
  \bibfield  {author} {\bibinfo {author} {\bibfnamefont {P.~A.}\ \bibnamefont
  {Kollman}}, \bibinfo {author} {\bibfnamefont {R.}~\bibnamefont {Dixon}},
  \bibinfo {author} {\bibfnamefont {W.}~\bibnamefont {Cornell}}, \bibinfo
  {author} {\bibfnamefont {T.}~\bibnamefont {Fox}}, \bibinfo {author}
  {\bibfnamefont {C.}~\bibnamefont {Chipot}}, \ and\ \bibinfo {author}
  {\bibfnamefont {A.}~\bibnamefont {Pohorille}},\ }\enquote {\bibinfo {title}
  {Computer simulations of biological systems},}\ \ (\bibinfo  {publisher}
  {Escom, Netherlands},\ \bibinfo {year} {1997})\ Chap.\ \bibinfo {chapter}
  {The development/application of a `minimalist' organic/biochemical molecular
  mechanic force field using a combination of ab initio calculations and
  experimental data}, pp.\ \bibinfo {pages} {83--96}
\bibitem [{\citenamefont {Yoda}, \citenamefont {Sugita},\ and\ \citenamefont
  {Okamoto}(2004{\natexlab{a}})}]{YSO1}%
  \BibitemOpen
  \bibfield  {author} {\bibinfo {author} {\bibfnamefont {T.}~\bibnamefont
  {Yoda}}, \bibinfo {author} {\bibfnamefont {Y.}~\bibnamefont {Sugita}}, \ and\
  \bibinfo {author} {\bibfnamefont {Y.}~\bibnamefont {Okamoto}},\ }\bibfield
  {title} {\enquote {\bibinfo {title} {Comparisons of force fields for proteins
  by generalized-ensemble simulations},}\ }\href@noop {} {\bibfield  {journal}
  {\bibinfo  {journal} {Chem. Phys. Lett.}\ }\textbf {\bibinfo {volume}
  {386}},\ \bibinfo {pages} {460--467} (\bibinfo {year} {2004}{\natexlab{a}})}
\bibitem [{\citenamefont {Yoda}, \citenamefont {Sugita},\ and\ \citenamefont
  {Okamoto}(2004{\natexlab{b}})}]{YSO2}%
  \BibitemOpen
  \bibfield  {author} {\bibinfo {author} {\bibfnamefont {T.}~\bibnamefont
  {Yoda}}, \bibinfo {author} {\bibfnamefont {Y.}~\bibnamefont {Sugita}}, \ and\
  \bibinfo {author} {\bibfnamefont {Y.}~\bibnamefont {Okamoto}},\ }\bibfield
  {title} {\enquote {\bibinfo {title} {Secondary-structure preferences of force
  fields for proteins evaluated by generalized-ensemble simulations},}\
  }\href@noop {} {\bibfield  {journal} {\bibinfo  {journal} {Chem. Phys.}\
  }\textbf {\bibinfo {volume} {307}},\ \bibinfo {pages} {269--283} (\bibinfo
  {year} {2004}{\natexlab{b}})}
\bibitem [{\citenamefont {Sakae}\ and\ \citenamefont {Okamoto}(2003)}]{SO1}%
  \BibitemOpen
  \bibfield  {author} {\bibinfo {author} {\bibfnamefont {Y.}~\bibnamefont
  {Sakae}}\ and\ \bibinfo {author} {\bibfnamefont {Y.}~\bibnamefont
  {Okamoto}},\ }\bibfield  {title} {\enquote {\bibinfo {title} {Optimization of
  protein force-field parameters with the protein data bank},}\ }\href@noop {}
  {\bibfield  {journal} {\bibinfo  {journal} {Chem. Phys. Lett.}\ }\textbf
  {\bibinfo {volume} {382}},\ \bibinfo {pages} {626--636} (\bibinfo {year}
  {2003})}
\bibitem [{\citenamefont {Sakae}\ and\ \citenamefont
  {Okamoto}(2004{\natexlab{a}})}]{SO2}%
  \BibitemOpen
  \bibfield  {author} {\bibinfo {author} {\bibfnamefont {Y.}~\bibnamefont
  {Sakae}}\ and\ \bibinfo {author} {\bibfnamefont {Y.}~\bibnamefont
  {Okamoto}},\ }\bibfield  {title} {\enquote {\bibinfo {title} {Protein
  force-field parameters optimized with the protein data bank. i. force-field
  optimizations},}\ }\href@noop {} {\bibfield  {journal} {\bibinfo  {journal}
  {J. Theo. Comput. Chem.}\ }\textbf {\bibinfo {volume} {3}},\ \bibinfo {pages}
  {339--358} (\bibinfo {year} {2004}{\natexlab{a}})}
\bibitem [{\citenamefont {Sakae}\ and\ \citenamefont
  {Okamoto}(2004{\natexlab{b}})}]{SO3}%
  \BibitemOpen
  \bibfield  {author} {\bibinfo {author} {\bibfnamefont {Y.}~\bibnamefont
  {Sakae}}\ and\ \bibinfo {author} {\bibfnamefont {Y.}~\bibnamefont
  {Okamoto}},\ }\bibfield  {title} {\enquote {\bibinfo {title} {Protein
  force-field parameters optimized with the protein data bank. ii. comparisons
  of force fields by folding simulations of short peptides},}\ }\href@noop {}
  {\bibfield  {journal} {\bibinfo  {journal} {J. Theo. Comput. Chem.}\ }\textbf
  {\bibinfo {volume} {3}},\ \bibinfo {pages} {359--378} (\bibinfo {year}
  {2004}{\natexlab{b}})}
\bibitem [{\citenamefont {Simmerling}, \citenamefont {Strockbine},\ and\
  \citenamefont {Roitberg}(2002)}]{Carlos}%
  \BibitemOpen
  \bibfield  {author} {\bibinfo {author} {\bibfnamefont {C.}~\bibnamefont
  {Simmerling}}, \bibinfo {author} {\bibfnamefont {B.}~\bibnamefont
  {Strockbine}}, \ and\ \bibinfo {author} {\bibfnamefont {A.~E.}\ \bibnamefont
  {Roitberg}},\ }\bibfield  {title} {\enquote {\bibinfo {title} {All-atom
  structure prediction and folding simulations of a stable protein},}\
  }\href@noop {} {\bibfield  {journal} {\bibinfo  {journal} {J. Am. Chem.
  Soc.}\ }\textbf {\bibinfo {volume} {124}},\ \bibinfo {pages} {11258--11259}
  (\bibinfo {year} {2002})}
\bibitem [{\citenamefont {Duan}\ \emph
  {et~al.}(2003{\natexlab{b}})\citenamefont {Duan}, \citenamefont {Wu},
  \citenamefont {Chowdhury}, \citenamefont {Lee}, \citenamefont {Xiong},
  \citenamefont {Zhang}, \citenamefont {Yang}, \citenamefont {Cieplak},
  \citenamefont {Luo}, \citenamefont {Lee}, \citenamefont {Caldwell},
  \citenamefont {Wang},\ and\ \citenamefont {Kollman}}]{Duan}%
  \BibitemOpen
  \bibfield  {author} {\bibinfo {author} {\bibfnamefont {Y.}~\bibnamefont
  {Duan}}, \bibinfo {author} {\bibfnamefont {C.}~\bibnamefont {Wu}}, \bibinfo
  {author} {\bibfnamefont {S.}~\bibnamefont {Chowdhury}}, \bibinfo {author}
  {\bibfnamefont {M.~C.}\ \bibnamefont {Lee}}, \bibinfo {author} {\bibfnamefont
  {G.}~\bibnamefont {Xiong}}, \bibinfo {author} {\bibfnamefont
  {W.}~\bibnamefont {Zhang}}, \bibinfo {author} {\bibfnamefont
  {R.}~\bibnamefont {Yang}}, \bibinfo {author} {\bibfnamefont {P.}~\bibnamefont
  {Cieplak}}, \bibinfo {author} {\bibfnamefont {R.}~\bibnamefont {Luo}},
  \bibinfo {author} {\bibfnamefont {T.}~\bibnamefont {Lee}}, \bibinfo {author}
  {\bibfnamefont {J.}~\bibnamefont {Caldwell}}, \bibinfo {author}
  {\bibfnamefont {J.}~\bibnamefont {Wang}}, \ and\ \bibinfo {author}
  {\bibfnamefont {P.}~\bibnamefont {Kollman}},\ }\bibfield  {title} {\enquote
  {\bibinfo {title} {A point-charge force field for molecular mechanics
  simulations of proteins based on condensed-phase quantum mechanical
  calculations},}\ }\href@noop {} {\bibfield  {journal} {\bibinfo  {journal}
  {J. Comput. Chem.}\ }\textbf {\bibinfo {volume} {24}},\ \bibinfo {pages}
  {1999--2012} (\bibinfo {year} {2003}{\natexlab{b}})}
\bibitem [{\citenamefont {Iwaoka}\ and\ \citenamefont {Tomoda}(2003)}]{IWA}%
  \BibitemOpen
  \bibfield  {author} {\bibinfo {author} {\bibfnamefont {M.}~\bibnamefont
  {Iwaoka}}\ and\ \bibinfo {author} {\bibfnamefont {S.}~\bibnamefont
  {Tomoda}},\ }\bibfield  {title} {\enquote {\bibinfo {title} {The saap force
  field. a simple approach to a new all-atom protein force field by using
  single amino acid potential (saap) functions in various solvents},}\
  }\href@noop {} {\bibfield  {journal} {\bibinfo  {journal} {J. Comput. Chem.}\
  }\textbf {\bibinfo {volume} {24}},\ \bibinfo {pages} {1192--1200} (\bibinfo
  {year} {2003})}
\bibitem [{\citenamefont {MacKerell~Jr}, \citenamefont {Feig},\ and\
  \citenamefont {Brooks~III}()}]{MFB}%
  \BibitemOpen
  \bibfield  {author} {\bibinfo {author} {\bibfnamefont {A.~D.}\ \bibnamefont
  {MacKerell~Jr}}, \bibinfo {author} {\bibfnamefont {M.}~\bibnamefont {Feig}},
  \ and\ \bibinfo {author} {\bibfnamefont {C.~L.}\ \bibnamefont {Brooks~III}},\
  }\bibfield  {title} {\enquote {\bibinfo {title} {Extending the treatment of
  backbone energetics in protein force fields: Limitations of gas-phase quantum
  mechanics in reproducing protein conformational distributions in molecular
  dynamics simulations},}\ }\href@noop {} {\bibinfo  {journal} {J Comput.
  Chem.}\ }
\bibitem [{\citenamefont {Kamiya}\ \emph {et~al.}(2005)\citenamefont {Kamiya},
  \citenamefont {Watanabe}, \citenamefont {Ono},\ and\ \citenamefont
  {Higo}}]{Kamiya}%
  \BibitemOpen
\bibfield  {journal} {  }\bibfield  {author} {\bibinfo {author} {\bibfnamefont
  {N.}~\bibnamefont {Kamiya}}, \bibinfo {author} {\bibfnamefont
  {Y.}~\bibnamefont {Watanabe}}, \bibinfo {author} {\bibfnamefont
  {S.}~\bibnamefont {Ono}}, \ and\ \bibinfo {author} {\bibfnamefont
  {J.}~\bibnamefont {Higo}},\ }\bibfield  {title} {\enquote {\bibinfo {title}
  {Amber-based hybrid force field for conformational sampling of
  polypeptides},}\ }\href@noop {} {\bibfield  {journal} {\bibinfo  {journal}
  {Chem. Phys. Lett.}\ }\textbf {\bibinfo {volume} {401}},\ \bibinfo {pages}
  {312--317} (\bibinfo {year} {2005})}
\bibitem [{\citenamefont {Best}\ and\ \citenamefont
  {Hummer}(2009)}]{Hummer2009}%
  \BibitemOpen
  \bibfield  {author} {\bibinfo {author} {\bibfnamefont {R.~B.}\ \bibnamefont
  {Best}}\ and\ \bibinfo {author} {\bibfnamefont {G.}~\bibnamefont {Hummer}},\
  }\bibfield  {title} {\enquote {\bibinfo {title} {Optimized molecular dynamics
  force field applied to the helix-coil transition of polypeptides},}\
  }\href@noop {} {\bibfield  {journal} {\bibinfo  {journal} {J. Phys. Chem. B}\
  }\textbf {\bibinfo {volume} {113}},\ \bibinfo {pages} {9004--9015} (\bibinfo
  {year} {2009})}
\bibitem [{\citenamefont {Mittal}\ and\ \citenamefont
  {Best}(2010)}]{Best_folding}%
  \BibitemOpen
  \bibfield  {author} {\bibinfo {author} {\bibfnamefont {J.}~\bibnamefont
  {Mittal}}\ and\ \bibinfo {author} {\bibfnamefont {R.~B.}\ \bibnamefont
  {Best}},\ }\bibfield  {title} {\enquote {\bibinfo {title} {Tackling
  force-field bias in protein folding simulations: Folding of villin hp35 and
  pin ww domains in explicit water},}\ }\href@noop {} {\bibfield  {journal}
  {\bibinfo  {journal} {Biophys. J.}\ }\textbf {\bibinfo {volume} {99}},\
  \bibinfo {pages} {L26--L28} (\bibinfo {year} {2010})}
\bibitem [{\citenamefont {Sakae}\ and\ \citenamefont {Okamoto}(2006)}]{SO4}%
  \BibitemOpen
  \bibfield  {author} {\bibinfo {author} {\bibfnamefont {Y.}~\bibnamefont
  {Sakae}}\ and\ \bibinfo {author} {\bibfnamefont {Y.}~\bibnamefont
  {Okamoto}},\ }\bibfield  {title} {\enquote {\bibinfo {title}
  {Secondary-structure design of proteins by a backbone torsion energy},}\
  }\href@noop {} {\bibfield  {journal} {\bibinfo  {journal} {J. Phys. Soc.
  Jpn.}\ }\textbf {\bibinfo {volume} {75}} (\bibinfo {year} {2006})},\ \bibinfo
  {note} {054802 (9 pages)}
\bibitem [{\citenamefont {Sakae}\ and\ \citenamefont
  {Okamoto}(2010{\natexlab{a}})}]{SO5}%
  \BibitemOpen
  \bibfield  {author} {\bibinfo {author} {\bibfnamefont {Y.}~\bibnamefont
  {Sakae}}\ and\ \bibinfo {author} {\bibfnamefont {Y.}~\bibnamefont
  {Okamoto}},\ }\bibfield  {title} {\enquote {\bibinfo {title} {Controlling the
  secondary-structure-forming tendencies of proteins by a backbone
  torsion-energy term},}\ }\href@noop {} {\bibfield  {journal} {\bibinfo
  {journal} {Mol. Sim.}\ }\textbf {\bibinfo {volume} {36}},\ \bibinfo {pages}
  {138--158} (\bibinfo {year} {2010}{\natexlab{a}})}
\bibitem [{\citenamefont {Ramachandran}\ and\ \citenamefont
  {Sasisekharan}(1968)}]{Rama_Sasi}%
  \BibitemOpen
  \bibfield  {author} {\bibinfo {author} {\bibfnamefont {G.~N.}\ \bibnamefont
  {Ramachandran}}\ and\ \bibinfo {author} {\bibfnamefont {V.}~\bibnamefont
  {Sasisekharan}},\ }\bibfield  {title} {\enquote {\bibinfo {title}
  {Conformation of polypeptides and proteins},}\ }\href@noop {} {\bibfield
  {journal} {\bibinfo  {journal} {Adv. Protein Chem.}\ }\textbf {\bibinfo
  {volume} {23}},\ \bibinfo {pages} {283--438} (\bibinfo {year} {1968})}
\bibitem [{\citenamefont {Sakae}\ and\ \citenamefont
  {Okamoto}(2010{\natexlab{b}})}]{SO6}%
  \BibitemOpen
  \bibfield  {author} {\bibinfo {author} {\bibfnamefont {Y.}~\bibnamefont
  {Sakae}}\ and\ \bibinfo {author} {\bibfnamefont {Y.}~\bibnamefont
  {Okamoto}},\ }\bibfield  {title} {\enquote {\bibinfo {title} {Determination
  method of the balance of the secondary-structure-forming tendencies of force
  fields},}\ }\href@noop {} {\bibfield  {journal} {\bibinfo  {journal} {Mol.
  Sim.}\ }\textbf {\bibinfo {volume} {36}},\ \bibinfo {pages} {159--165}
  (\bibinfo {year} {2010}{\natexlab{b}})}
\bibitem [{\citenamefont {Sakae}\ and\ \citenamefont
  {Okamoto}(2010{\natexlab{c}})}]{SO7}%
  \BibitemOpen
  \bibfield  {author} {\bibinfo {author} {\bibfnamefont {Y.}~\bibnamefont
  {Sakae}}\ and\ \bibinfo {author} {\bibfnamefont {Y.}~\bibnamefont
  {Okamoto}},\ }\bibfield  {title} {\enquote {\bibinfo {title} {Optimisation of
  opls-ua force-field parameters for protein systems using protein data
  bank},}\ }\href@noop {} {\bibfield  {journal} {\bibinfo  {journal} {Mol.
  Sim.}\ }\textbf {\bibinfo {volume} {36}},\ \bibinfo {pages} {1148--1156}
  (\bibinfo {year} {2010}{\natexlab{c}})}
\bibitem [{\citenamefont {Noguchi}\ \emph {et~al.}(1997)\citenamefont
  {Noguchi}, \citenamefont {Onizuka}, \citenamefont {Akiyama},\ and\
  \citenamefont {Saito}}]{REPRDB}%
  \BibitemOpen
  \bibfield  {author} {\bibinfo {author} {\bibfnamefont {T.}~\bibnamefont
  {Noguchi}}, \bibinfo {author} {\bibfnamefont {K.}~\bibnamefont {Onizuka}},
  \bibinfo {author} {\bibfnamefont {Y.}~\bibnamefont {Akiyama}}, \ and\
  \bibinfo {author} {\bibfnamefont {M.}~\bibnamefont {Saito}},\ }\bibfield
  {title} {\enquote {\bibinfo {title} {Pdb-reprdb: A database of representative
  protein chains in pdb (protein data bank)},}\ }in\ \href@noop {} {\emph
  {\bibinfo {booktitle} {Proc. of the Fifth International Conference on
  Intelligent Systems for Molecular Biology}}}\ (\bibinfo  {publisher} {AAAI
  press},\ \bibinfo {address} {Menlo Park, CA},\ \bibinfo {year} {1997})
\bibitem [{\citenamefont {Case}\ \emph {et~al.}(2005)\citenamefont {Case},
  \citenamefont {Cheatham}, \citenamefont {Darden}, \citenamefont {Gohlke},
  \citenamefont {Luo}, \citenamefont {Merz}, \citenamefont {Onufriev},
  \citenamefont {Simmerling}, \citenamefont {Wang},\ and\ \citenamefont
  {Woods}}]{amber_prog11}%
  \BibitemOpen
  \bibfield  {author} {\bibinfo {author} {\bibfnamefont {D.~A.}\ \bibnamefont
  {Case}}, \bibinfo {author} {\bibfnamefont {T.}~\bibnamefont {Cheatham}},
  \bibinfo {author} {\bibfnamefont {T.}~\bibnamefont {Darden}}, \bibinfo
  {author} {\bibfnamefont {H.}~\bibnamefont {Gohlke}}, \bibinfo {author}
  {\bibfnamefont {R.}~\bibnamefont {Luo}}, \bibinfo {author} {\bibfnamefont
  {K.~M.}\ \bibnamefont {Merz}, \bibfnamefont {Jr.}}, \bibinfo {author}
  {\bibfnamefont {A.}~\bibnamefont {Onufriev}}, \bibinfo {author}
  {\bibfnamefont {C.}~\bibnamefont {Simmerling}}, \bibinfo {author}
  {\bibfnamefont {B.}~\bibnamefont {Wang}}, \ and\ \bibinfo {author}
  {\bibfnamefont {R.}~\bibnamefont {Woods}},\ }\bibfield  {title} {\enquote
  {\bibinfo {title} {The amber biomolecular simulation programs},}\ }\href@noop
  {} {\bibfield  {journal} {\bibinfo  {journal} {J. Computat. Chem.}\ }\textbf
  {\bibinfo {volume} {26}},\ \bibinfo {pages} {1668--1688} (\bibinfo {year}
  {2005})}
\bibitem [{\citenamefont {Onufriev}, \citenamefont {Bashford},\ and\
  \citenamefont {Case}(2004)}]{gbsa_igb5}%
  \BibitemOpen
  \bibfield  {author} {\bibinfo {author} {\bibfnamefont {A.}~\bibnamefont
  {Onufriev}}, \bibinfo {author} {\bibfnamefont {D.}~\bibnamefont {Bashford}},
  \ and\ \bibinfo {author} {\bibfnamefont {D.~A.}\ \bibnamefont {Case}},\
  }\bibfield  {title} {\enquote {\bibinfo {title} {Exploring protein native
  states and large-scale conformational changes with a modified generalized
  born model},}\ }\href@noop {} {\bibfield  {journal} {\bibinfo  {journal}
  {Proteins}\ }\textbf {\bibinfo {volume} {55}},\ \bibinfo {pages} {383--394}
  (\bibinfo {year} {2004})}
\bibitem [{\citenamefont {Weiser}, \citenamefont {Shenkin},\ and\ \citenamefont
  {Still}(1999)}]{gbsa_sa1}%
  \BibitemOpen
  \bibfield  {author} {\bibinfo {author} {\bibfnamefont {J.}~\bibnamefont
  {Weiser}}, \bibinfo {author} {\bibfnamefont {P.~S.}\ \bibnamefont {Shenkin}},
  \ and\ \bibinfo {author} {\bibfnamefont {W.~C.}\ \bibnamefont {Still}},\
  }\bibfield  {title} {\enquote {\bibinfo {title} {Approximate atomic surfaces
  from linear combinations of pairwise overlaps (lcpo)},}\ }\href@noop {}
  {\bibfield  {journal} {\bibinfo  {journal} {J. Comput. Chem.}\ }\textbf
  {\bibinfo {volume} {20}},\ \bibinfo {pages} {217--230} (\bibinfo {year}
  {1999})}
\bibitem [{\citenamefont {Honda}, \citenamefont {Kobayashi},\ and\
  \citenamefont {Munekata}(2000)}]{gpep3}%
  \BibitemOpen
  \bibfield  {author} {\bibinfo {author} {\bibfnamefont {S.}~\bibnamefont
  {Honda}}, \bibinfo {author} {\bibfnamefont {N.}~\bibnamefont {Kobayashi}}, \
  and\ \bibinfo {author} {\bibfnamefont {E.}~\bibnamefont {Munekata}},\
  }\bibfield  {title} {\enquote {\bibinfo {title} {Thermodynamics of a
  β-hairpin structure: evidence for cooperative formation of folding
  nucleus},}\ }\href@noop {} {\bibfield  {journal} {\bibinfo  {journal} {J.
  Mol. Biol.}\ }\textbf {\bibinfo {volume} {295}},\ \bibinfo {pages} {269--278}
  (\bibinfo {year} {2000})}
\bibitem [{\citenamefont {Shoemaker}\ \emph {et~al.}(1985)\citenamefont
  {Shoemaker}, \citenamefont {Kim}, \citenamefont {Brems}, \citenamefont
  {Marqusee}, \citenamefont {York}, \citenamefont {Chaiken}, \citenamefont
  {Stewart},\ and\ \citenamefont {Baldwin}}]{buzz1}%
  \BibitemOpen
  \bibfield  {author} {\bibinfo {author} {\bibfnamefont {K.~R.}\ \bibnamefont
  {Shoemaker}}, \bibinfo {author} {\bibfnamefont {P.~S.}\ \bibnamefont {Kim}},
  \bibinfo {author} {\bibfnamefont {D.~N.}\ \bibnamefont {Brems}}, \bibinfo
  {author} {\bibfnamefont {S.}~\bibnamefont {Marqusee}}, \bibinfo {author}
  {\bibfnamefont {E.~J.}\ \bibnamefont {York}}, \bibinfo {author}
  {\bibfnamefont {I.~M.}\ \bibnamefont {Chaiken}}, \bibinfo {author}
  {\bibfnamefont {J.~M.}\ \bibnamefont {Stewart}}, \ and\ \bibinfo {author}
  {\bibfnamefont {R.~L.}\ \bibnamefont {Baldwin}},\ }\bibfield  {title}
  {\enquote {\bibinfo {title} {Nature of the charged-group effect on the
  stability of the c-peptide helix},}\ }\href@noop {} {\bibfield  {journal}
  {\bibinfo  {journal} {Proc. Natl. Acad. Sci. U.S.A.}\ }\textbf {\bibinfo
  {volume} {82}},\ \bibinfo {pages} {2349--2353} (\bibinfo {year} {1985})}
\bibitem [{\citenamefont {Osterhout~Jr.}\ \emph {et~al.}(1989)\citenamefont
  {Osterhout~Jr.}, \citenamefont {Baldwin}, \citenamefont {York}, \citenamefont
  {Stewart}, \citenamefont {Dyson},\ and\ \citenamefont {Wright}}]{buzz2}%
  \BibitemOpen
  \bibfield  {author} {\bibinfo {author} {\bibfnamefont {J.~J.}\ \bibnamefont
  {Osterhout~Jr.}}, \bibinfo {author} {\bibfnamefont {R.~L.}\ \bibnamefont
  {Baldwin}}, \bibinfo {author} {\bibfnamefont {E.~J.}\ \bibnamefont {York}},
  \bibinfo {author} {\bibfnamefont {J.~M.}\ \bibnamefont {Stewart}}, \bibinfo
  {author} {\bibfnamefont {H.~J.}\ \bibnamefont {Dyson}}, \ and\ \bibinfo
  {author} {\bibfnamefont {P.~E.}\ \bibnamefont {Wright}},\ }\bibfield  {title}
  {\enquote {\bibinfo {title} {1h nmr studies of the solution conformations of
  an analogue of the c-peptide of ribonuclease a},}\ }\href@noop {} {\bibfield
  {journal} {\bibinfo  {journal} {Biochemistry}\ }\textbf {\bibinfo {volume}
  {28}},\ \bibinfo {pages} {7059--7064} (\bibinfo {year} {1989})}
\bibitem [{\citenamefont {Blanco}, \citenamefont {Rivas},\ and\ \citenamefont
  {Serrano}(1994)}]{gpep1}%
  \BibitemOpen
  \bibfield  {author} {\bibinfo {author} {\bibfnamefont {F.~J.}\ \bibnamefont
  {Blanco}}, \bibinfo {author} {\bibfnamefont {G.}~\bibnamefont {Rivas}}, \
  and\ \bibinfo {author} {\bibfnamefont {L.}~\bibnamefont {Serrano}},\
  }\bibfield  {title} {\enquote {\bibinfo {title} {A short linear peptide that
  folds into a native stable bold beta-hairpin in aqueous solution},}\
  }\href@noop {} {\bibfield  {journal} {\bibinfo  {journal} {Nature Struct.
  Biol.}\ }\textbf {\bibinfo {volume} {1}},\ \bibinfo {pages} {584--590}
  (\bibinfo {year} {1994})}
\bibitem [{\citenamefont {Kobayashi}\ \emph {et~al.}(1995)\citenamefont
  {Kobayashi}, \citenamefont {Honda}, \citenamefont {Yoshii}, \citenamefont
  {Uedaira},\ and\ \citenamefont {Munekata}}]{gpep2}%
  \BibitemOpen
  \bibfield  {author} {\bibinfo {author} {\bibfnamefont {N.}~\bibnamefont
  {Kobayashi}}, \bibinfo {author} {\bibfnamefont {S.}~\bibnamefont {Honda}},
  \bibinfo {author} {\bibfnamefont {H.}~\bibnamefont {Yoshii}}, \bibinfo
  {author} {\bibfnamefont {H.}~\bibnamefont {Uedaira}}, \ and\ \bibinfo
  {author} {\bibfnamefont {E.}~\bibnamefont {Munekata}},\ }\bibfield  {title}
  {\enquote {\bibinfo {title} {Complement assembly of two fragments of the
  streptococcal protein g b1 domain in aqueous solution},}\ }\href@noop {}
  {\bibfield  {journal} {\bibinfo  {journal} {FEBS Lett.}\ }\textbf {\bibinfo
  {volume} {366}},\ \bibinfo {pages} {99--103} (\bibinfo {year} {1995})}
\bibitem [{\citenamefont {Sugita}\ and\ \citenamefont {Okamoto}(1999)}]{REMD}%
  \BibitemOpen
  \bibfield  {author} {\bibinfo {author} {\bibfnamefont {Y.}~\bibnamefont
  {Sugita}}\ and\ \bibinfo {author} {\bibfnamefont {Y.}~\bibnamefont
  {Okamoto}},\ }\bibfield  {title} {\enquote {\bibinfo {title}
  {Replica-exchange molecular dynamics method for protein folding},}\
  }\href@noop {} {\bibfield  {journal} {\bibinfo  {journal} {Chem. Phys.
  Lett.}\ }\textbf {\bibinfo {volume} {314}},\ \bibinfo {pages} {141--151}
  (\bibinfo {year} {1999})}
\bibitem [{\citenamefont {Ryckaert}, \citenamefont {Ciccotti},\ and\
  \citenamefont {Berendsen}(1977)}]{SHAKE}%
  \BibitemOpen
  \bibfield  {author} {\bibinfo {author} {\bibfnamefont {J.-P.}\ \bibnamefont
  {Ryckaert}}, \bibinfo {author} {\bibfnamefont {G.}~\bibnamefont {Ciccotti}},
  \ and\ \bibinfo {author} {\bibfnamefont {H.~J.~C.}\ \bibnamefont
  {Berendsen}},\ }\bibfield  {title} {\enquote {\bibinfo {title} {Numerical
  integration of the cartesian equations of motion of a system with
  constraints: Moecular dynamics of n-alkanes},}\ }\href@noop {} {\bibfield
  {journal} {\bibinfo  {journal} {J. Comput. Phys.}\ }\textbf {\bibinfo
  {volume} {23}},\ \bibinfo {pages} {327--341} (\bibinfo {year} {1977})}
\bibitem [{\citenamefont {Kabsch}\ and\ \citenamefont {Sander}(1983)}]{DSSP}%
  \BibitemOpen
  \bibfield  {author} {\bibinfo {author} {\bibfnamefont {W.}~\bibnamefont
  {Kabsch}}\ and\ \bibinfo {author} {\bibfnamefont {C.}~\bibnamefont
  {Sander}},\ }\bibfield  {title} {\enquote {\bibinfo {title} {Dictionary of
  protein secondary structure: Pattern recognition of hydrogen-bonded and
  geometrical features},}\ }\href@noop {} {\bibfield  {journal} {\bibinfo
  {journal} {Biopolymers}\ }\textbf {\bibinfo {volume} {22}},\ \bibinfo {pages}
  {2577--2637} (\bibinfo {year} {1983})}
\end{thebibliography}

%


~~\\
~~\\
~~\\
~~\\
~~\\
~~\\
~~\\
~~\\
~~\\
~~\\

\newpage


\begin{table}
\caption{Torsion-energy parameters ($V_n$ and $\gamma_n$)
for the main-chain dihedral angles $\psi$ and $\psi'$ 
in Eq.~(\ref{ene_torsion1})  
for the original AMBER ff94, ff96, ff99, ff99SB, and ff03 force
fields. 
The values are common among the
amino-acid residues for each force field.
Only the parameters for non-zero $V_n$ are listed.}
\label{table-org-torsion}
\vspace{0.3cm}
\begin{tabular}{lcccccc} \hline
force field  &   & $\psi$ (N-C$_{\alpha}$-C-N) &  &   & $\psi'$ (C$_{\beta}$-C$_{\alpha}$-C-N) &            \\ \hline
  & $n$ & $\displaystyle{V_n/2}$ & $\gamma_n$ & ~~~~~~~ $n$ & $\displaystyle{V_n/2}$  & $\gamma_n$ \\ \hline
ff94~~~~~   &  1  & 0.75 & $\pi$ &  ~~~~~~~ 2  & 0.07 &  0    \\
            &  2  & 1.35 & $\pi$ &  ~~~~~~~ 4  & 0.10 &  0    \\
            &  4  & 0.40 & $\pi$ &  ~~~~~~~    &      &         \\ \hline
ff96~~~~~   &  1  & 0.85 &   0   &  ~~~~~~~ 2  & 0.07 &  0    \\ 
            &  2  & 0.30 & $\pi$ &  ~~~~~~~ 4  & 0.10 &  0    \\ \hline
ff99~~~~~   &  1  & 1.70 & $\pi$ &  ~~~~~~~ 2  & 0.07 &  0    \\ 
            &  2  & 2.00 & $\pi$ &  ~~~~~~~ 4  & 0.10 &  0    \\ \hline
ff99SB~~~   &  1  & 0.45 & $\pi$ &  ~~~~~~~ 1  & 0.20 &  0    \\ 
            &  2  & 1.58 & $\pi$ &  ~~~~~~~ 2  & 0.20 &  0    \\ 
            &  3  & 0.55 & $\pi$ &  ~~~~~~~ 3  & 0.40 &  0    \\ \hline
ff03~~~~~   &  1  & 0.6839 & $\pi$ &  ~~~~~~~ 1  & 0.7784 &  $\pi$    \\ 
            &  2  & 1.4537 & $\pi$ &  ~~~~~~~ 2  & 0.0657 &  $\pi$    \\ 
            &  3  & 0.4615 & $\pi$ &  ~~~~~~~ 3  & 0.0560 &  0    \\ \hline

\end{tabular}
\end{table}

~~\\
~~\\
~~\\
~~\\
~~\\
~~\\
~~\\
~~\\
~~\\
~~\\
~~\\
~~\\

   

\begin{table}
\caption{100 proteins used in the optimization of force-field parameters.}
\label{PDB_list}
\vspace{0.3cm}
\begin{tabular}{lclclclcl} \hline
fold            &~~ PDB ID ~~&  chain ~&~~ PDB ID  ~~& chain ~&~~ PDB ID  ~~& chain ~&~~ PDB ID ~~& chain  \\ \hline
all $\alpha$    & 1DLW &  A  & 1N1J &  B  & 1U84 &  A  & 1HBK &  A  \\
                & 1TX4 &  A  & 1V54 &  E  & 1SK7 &  A  & 1TQG &  A  \\
                & 1V74 &  B  & 1DVO &  A  & 1HFE &  S  & 1J0P &  A  \\
                & 1Y02 &  A71-114  & 1IJY &  A  & 1I2T &  A  & 1G8E &  A  \\
                & 1VKE &  C  & 1FS1 & A109-149  &  1D9C  &  A  & 1AIL & A  \\
                & 1Q5Z &  A  & 1T8K &  A  & 1OR7 &  C  & 1NG6 & A    \\
                & 1C75 &  A  & 2LIS &  A  & 1NH2 &  B  & 1Q2H &  A  \\
                & 1NKP &  A  &      &     &      &     &      &     \\
 all $\beta$    & 1XAK &  A  & 1T2W &  A  & 1GMU &  C1-70 & 1AYO &  A  \\
                & 1PK6 &  A  & 1OFS &  B  & 1BEH &  A  & 1JO8 &  A  \\
                & 1UXZ &  A  & 1UB4 &  C  & 1LGP &  A  & 1CQY &  A  \\
                & 1PM4 &  A  & 1OU8 &  A  & 1V76 &  A  & 1R6J &  A  \\
                & 1OA8 &  D  & 1IFG &  A  &      &     &      &     \\
$\alpha / \beta$& 1IO0 &  A  & 1U7P &  A  & 1JKE &  C  & 1MXI &  A  \\
                & 1LY1 &  A  & 1NRZ &  A  & 1IM5 &  A  & 1VC1 &  A  \\
                & 1OGD &  A  & 1IIB &  A  & 1PYO &  D  & 1MUG &  A  \\
                & 1H75 &  A  & 1K66 &  A  & 1COZ &  A  & 1D4O &  A  \\
$\alpha+\beta$  & 1VCC &  A  & 1PP0 &  B  & 1PZ4 &  A  & 1TU1 &  A  \\
                & 1Q2Y &  A  & 1M4J &  A  & 1N9L &  A  & 1LQV &  B  \\
                & 1A3A &  A  & 1K2E &  A  & 1TT8 &  A  & 1HUF &  A  \\
                & 1SXR &  A  & 1CYO &  A  & 1ID0 &  A  & 1UCD &  A  \\
                & 1F46 &  B  & 1KPF &  A  & 1BYR &  A  & 1Y60 &  D  \\
                & 1SEI &  A  & 1RL6 &  A  & 1WM3 &  A  & 1FTH &  A  \\
                & 1APY &  B  & 1N13 &  E  & 1LTS &  C  & 1UGI &  A  \\
                & 1MWP &  A  & 1PCF &  A  & 1IHR &  B  & 1H6H &  A  \\ \hline
\end{tabular}
\end{table}


\begin{table}
\caption{Optimized $V_1/2$ parameters for the main-chain dihedral 
angles $\psi$ and $\psi'$ 
for the 19 amino-acid residues (except for proline) 
in Eq.~(\ref{ene_torsion3}). 
The rest of the parameters are taken to be the same as
in the original ff03 force field (see Table~\ref{table-org-torsion}).
The original amino-acid-independent values are also listed
for reference.}
\label{table-opt-parameters}
\vspace{0.3cm}
\begin{tabular}{lrcr} \hline
                 & $\psi$ (N-C$_{\alpha}$-C-N)  & ~~~ & $\psi'$ (C$_{\beta}$-C$_{\alpha}$-C-N)      \\ \hline
original ff03    & 0.6839      && 0.7784        \\ 
Ala              & 0.122      && 0.150        \\ 
Arg              & 0.409      && 0.200        \\ 
Asn              & $-0.074$    && $-0.162$       \\ 
Asp              & $-0.137$    && 0.182        \\ 
Cys              & 0.361      && 0.089        \\ 
Gln              & 0.144      && $-0.024$       \\ 
Glu              & 0.180      && 0.152        \\ 
Gly              & 0.258      && $---$         \\ 
His              & 0.020      && 0.237        \\ 
Ile              & 0.643      && 0.194        \\ 
Leu              & 0.382      && 0.257        \\ 
Lys              & 0.222      && 0.042        \\ 
Met              & 0.141      && 0.346        \\ 
Phe              & $-0.010$     && 0.553        \\ 
Ser              & $-0.248$     && 0.475        \\ 
Thr              & 0.512      && 0.328        \\ 
Trp              & 0.027      && 0.477        \\ 
Tyr              & 0.082      && 0.652        \\ 
Val              & 0.142      && 0.590        \\ \hline
\end{tabular}
\end{table}


\end{document}